I

# European Aerosol Phenomenology - 8: Harmonised Source Apportionment of Organic Aerosol using 22 Year-long ACSM/AMS Datasets

Gang Chen[1], Francesco Canonaco[1,2], Anna Tobler[1,2], Wenche Aas[3], Andres Alastuey[4], James Allan[5,6], Samira Atabakhsh[7], Minna Aurela[8], Urs Baltensperger[1], Aikaterini Bougiatioti[9], Joel F. De Brito[10], Darius Ceburnis[11], Benjamin Chazeau[1,12,13], Hasna Chebaicheb[10,14], Kaspar R. Daellenbach[1], Mikael Ehn[15], Imad El Haddad[1], Konstantinos Eleftheriadis[16], Olivier Favez[14], Harald Flentje[17], Anna Font[18*], Kirsten Fossum[11], Evelyn Freney[14,20], Maria Gini[16], David C Green[18,19], Liine Heikkinen[15**], Hartmut Herrmann[7], Athina-Cerise Kalogridis[16], Hannes Keernik[21,22], Radek Lhotka[23,24], Chunshui Lin[11], Chris Lunder[3], Marek Maasikmets[21], Manousos I. Manousakas[1], Nicolas Marchand[12], Cristina Marin[25,26], Luminita Marmureanu[25], Nikolaos Mihalopoulos[9], Griša Močnik[27,28], Jaroslaw Nęcki[29], Colin O'Dowd[11], Jurgita Ovadnevaite[11], Thomas Peter[30], Jean-Eudes Petit[31], Michael Pikridas[32], Stephen Matthew Platt[3], Petra Pokorná[23], Laurent Poulain[7], Max Priestman[18], Véronique Riffault[10], Matteo Rinaldi[33], Kazimierz Różański[29], Jaroslav Schwarz[23], Jean Sciare[32], Leïla Simon[14,31], Alicja Skiba[29], Jay G. Slowik[1], Yulia Sosedova[2], Iasonas Stavroulas[9,32], Katarzyna Styszko[34], Erik Teinemaa[21], Hilkka Timonen[8], Anja Tremper[18,19], Jeni Vasilescu[25], Marta Via[4,35], Petr Vodička[23], Alfred Wiedensohler[7], Olga Zografou[16], María Cruz Minguillón[4***], and André S.H. Prévôt[1***]

[1]Laboratory of Atmospheric Chemistry, Paul Scherrer Institute, 5232 Villigen, Switzerland
[2]Datalystica Ltd., Park innovAARE, 5234 Villigen, Switzerland
[3]NILU - Norwegian Institute for Air Research, 2007 Kjeller, Norway
[4]Institute of Environmental Assessment and Water Research (IDAEA), Spanish Council for Scientific Research (CSIC), Barcelona, 08034, Spain
[5]Department of Earth and Environmental Sciences, University of Manchester, Manchester, UK
[6]National Centre for Atmospheric Science, University of Manchester, Manchester, UK
[7]Department of Chemistry of the Atmosphere Leibniz Institute for Tropospheric Research, Permoser Straße 15, 04318, Leipzig, Germany
[8]Atmospheric Composition Research, Finnish Meteorological Institute, P.O. Box 503, 00101, Helsinki, Finland
[9]Institute for Environmental Research and Sustainable Development, National Observatory of Athens, Palaia Penteli, 15236, Athens, Greece
[10] IMT Nord Europe, Institut Mines-Télécom, Univ. Lille, Centre for Energy and Environment, 59000 Lille, France
[11]School of Physics, Ryan Institute's Centre for Climate and Air Pollution Studies, National University of Ireland Galway, University Road, Galway, H91 CF50, Ireland
[12]Aix Marseille Univ., CNRS, LCE, Marseille, France
[13]AtmoSud, Regional Network for Air Quality Monitoring of Provence-Alpes-Côte-d'Azur, Marseille, France
[14]Institut National de l'Environnement Industriel et des Risques, Parc Technologique ALATA, 60550, Verneuil en Halatte, France

[15] Institute for Atmospheric and Earth System Research (INAR) / Physics, University of Helsinki, Helsinki, Finland

[16] Environmental Radioactivity Laboratory, Institute of Nuclear & Radiological Sciences & Technology, Energy & Safety, N.C.S.R. "Demokritos", 15310 Athens, Greece

[17] Deutscher Wetterdienst, Meteorologisches Observatorium Hohenpeißenberg, 82383 Hohenpeißenberg, Germany

[18] MRC Centre for Environment and Health, Environmental Research Group, Imperial College London, 86 Wood Lane, London, W12 0BZ, UK

[19] HPRU in Environmental Exposures and Health, Imperial College London, UK

[20] Laboratoire de Météorologie Physique, UMR6016, Université Clermont Auvergne-CNRS, Aubière, France

[21] Air Quality and Climate Department, Estonian Environmental Research Centre (EERC), Marja 4D, Tallinn, Estonia

[22] Department of Software Science, Tallinn University of Technology, 19086 Tallinn, Estonia

[23] Institute of Chemical Process Fundamentals of the CAS, Rozvojová 135/1, 16502 Prague, Czech Republic

[24] Institute for Environmental Studies, Faculty of Science, Charles University, Benátská 2, 12801 Prague, Czech Republic

[25] National Institute of Research and Development for Optoelectronics INOE 2000, 77125 Magurele, Romania

[26] Department of Physics, Politehnica University of Bucharest, Bucharest, Romania

[27] Condensed Matter Physics Department, J. Stefan Institute, Ljubljana, Slovenia

[28] Center for Atmospheric Research, University of Nova Gorica, Ajdovščina, Slovenia

[29] AGH University of Science and Technology, Faculty of Physics and Applied Computer Science, Department of Applied Nuclear Physics, Kraków, Poland

[30] Institute for Atmospheric and Climate Sciences, ETH Zürich, Zürich, 8092, Switzerland

[31] Laboratoire des Sciences du Climat et de l'Environnement, UMR 8212, CEA/Orme des Merisiers, 91191 Gif-sur-Yvette, France

[32] Climate & Atmosphere Research Centre (CARE-C), The Cyprus Institute, Nicosia, 2121, Cyprus

[33] Institute of Atmospheric Sciences and Climate (ISAC), National Research Council (CNR), 40129 Bologna, Italy

[34] AGH University of Science and Technology, Faculty of Energy and Fuels, Department of Coal Chemistry and Environmental Sciences, Kraków, Poland

[35] Department of Applied Physics, University of Barcelona, Barcelona, 08028, Spain

*Now at: IMT Nord Europe, Institut Mines-Télécom, Univ. Lille, Centre for Energy and Environment, 59000 Lille, France

**Now at: Department of Environmental Science & Bolin Centre for Climate Research, Stockholm University, Stockholm, Sweden

***Correspondence to: María Cruz Minguillón (mariacruz.minguillon@idaea.csic.es) and André S. H. Prévôt (andre.prevot@psi.ch)

# Abstract

Organic aerosol (OA) is a key component to total submicron particulate matter ($PM_1$), and comprehensive knowledge of OA sources across Europe is crucial to mitigate $PM_1$ levels. Europe has a well-established air quality research infrastructure from which yearlong datasets using 21 aerosol chemical speciation monitors (ACSMs) and 1 aerosol mass spectrometer (AMS) were gathered during 2013-2019. It includes 9 non-urban and 13 urban sites. This study developed a state-of-the-art source apportionment protocol to analyse long-term OA mass spectrum data by applying the most advanced source apportionment strategies (i.e., rolling PMF, ME-2, and bootstrap). This harmonised protocol was followed strictly for all 22 datasets, making the source apportionment results more comparable. In addition, it enables the quantifications of the most common OA components such as hydrocarbon-like OA (HOA), biomass burning OA (BBOA), cooking-like OA (COA), more oxidised-oxygenated OA (MO-OOA), and less oxidised-oxygenated OA (LO-OOA). Other components such as coal combustion OA (CCOA), solid fuel OA (SFOA: mainly mixture of coal and peat combustion), cigarette smoke OA (CSOA), sea salt (mostly inorganic but part of the OA mass spectrum), coffee OA, and ship industry OA could also be separated at a few specific sites. Oxygenated OA (OOA) components make up most of the submicron OA mass (average = 71.1%, range from 43.7 to 100%). Solid fuel combustion-related OA components (i.e., BBOA, CCOA, and SFOA) are still considerable with in total 16.0% yearly contribution to the OA, yet mainly during winter months (21.4%). Overall, this comprehensive protocol works effectively across all sites governed by different sources and generates robust and consistent source apportionment results. Our work presents a comprehensive overview of OA sources in Europe with a unique combination of high time resolution (30-240 minutes) and long-

IV

term data coverage (9-36 months), providing essential information to improve/validate air quality, health impact, and climate models.

# 1 Introduction

Atmospheric aerosols are liquid or solid particles suspended in the atmosphere (Hinds, 1999), which cause serious adverse health effects, reduce visibility, and interact with ecosystems and climate (IPCC, 2021). Despite all efforts, Europe is still suffering from poor air quality. Specifically, European Environment Agency (EEA) (2021) reported that 2% of the reporting countries exceeded the annual $PM_{2.5}$ (atmospheric particulate matter (PM) with an aerodynamic diameter of 2.5 μm or less) limit value of EU legislation (25 μg/m$^3$) in 2019. Following the even stricter 2021 WHO $PM_{2.5}$ guidelines (5 μg/m$^3$), all reporting countries, except Estonia, were over the limit that year (European Environment Agency (EEA), 2021; World Health Organization, 2021). Therefore, it is more important than ever to mitigate the air pollution levels. Moreover, premature deaths attributed to long-term exposure to $PM_{2.5}$ in 27 EU member states reached 307,000 in 2019 (European Environment Agency (EEA), 2021). As one of the most significant aerosol components, organic aerosol (OA) gained extensive interest since it represents 20 to 90% mass of the total submicron aerosols (Crippa et al., 2014; Jimenez et al., 2009; Zhang et al., 2011, 2007). Importantly, a simple reduction of $PM_{2.5}$ or even OA might not be an effective strategy to mitigate the health impacts of aerosol because OA components from different sources have different toxicities (Daellenbach et al., 2020). In addition, various sources/compositions of OA can also have significant differences in climate forcing (Yang et al., 2018). Therefore, the comprehensive knowledge of the OA sources could provide more information for regional, global climate, or air quality models for emission inventories, parameterization, or validation.

Typically, OA sources are identified by using the receptor model, positive matrix factorisation (PMF) on data from the Aerodyne aerosol mass spectrometer (AMS; (Jayne et al., 2000)). Many studies have reported a broad spatial overview of OA sources, which provides the chemical composition of the bulk non-refractory $PM_1$ and, more recently, also $PM_{2.5}$ (Elser et al., 2016; Xu et al., 2017). Zhang et al. (2007) first reviewed the use of factor analysis to investigate OA sources in urban and rural/remote sites in the Northern Hemisphere. Jimenez et al. (2009) provided an overview of $PM_1$ chemical composition and OA sources worldwide (including eight European sites), while Ng et al. (2010) provided a big picture of OA sources over the Northern Hemisphere with various environments (including 43 AMS datasets). $PM_1$ chemical composition and OA sources in Central Europe, including Switzerland, Germany, Austria, France, and Liechtenstein, have been described more specifically in Lanz et al. (2010). Finally, an overview of the OA sources across Europe using 25 AMS datasets, combined with guidelines for source apportionment (SA) applications to AMS data, resulted from the work of (Crippa et al., 2014). However, the AMS requires labour-intensive maintenance, making it extremely difficult to run continuously over a long period. Most of the AMS datasets in previous studies only covered time periods up to a few months. Therefore, the seasonal variations of OA sources in Europe remain mostly unknown without consistent long-term SA studies. Along with the well-established European Aerosol, Clouds and Trace gases Research InfraStructure (ACTRIS, http://actris.eu), there are 1 AMS and 21 Aerodyne Aerosol Chemical Speciation Monitors (ACSM, (Fröhlich et al., 2013; Ng et al., 2011b)), which is a simpler and more robust version of AMS. With these data, it is possible to address a critical knowledge gap in the literature, that is to say, the seasonal, diurnal, and spatial variabilities of OA sources across Europe.

An important limitation of the conventional PMF applications for long-term datasets is that OA sources are assumed to be static over the entire sampling period, although the source profiles of OA have substantial seasonal variations (Canonaco et al., 2015; Crippa et al., 2014; Zhang et al., 2019). This study applies a novel SA technique, rolling PMF (Canonaco et al., 2021; Parworth et al., 2015), to account for the temporal variabilities of the factor profiles. Instead of running PMF for the whole dataset, this work applies a smaller time window (e.g., varying from 7-28 days), moving with daily steps over the entire dataset (typically around one year). This allows the model to adapt the factor profiles across different observational periods gradually. In addition, the rolling technique has been equipped with random input resampling and random variation of the constraints allowing for a quantitative estimate of the statistical and rotational uncertainties of the PMF solutions (Canonaco et al., 2021; Chen et al., 2021). Meanwhile, rolling PMF can facilitate the analysis while handling the long-term ACSM/AMS datasets by saving computational time compared to conventional seasonal bootstrap PMF.

The comparison of SA results from different sites is challenging, as they are not always equivalent due to the subjective decisions made by the analysts during the different PMF analysis steps, especially regarding the optimal number of factors, their identification, and validation. For that reason, a standardised protocol (Section 2.3) has been developed to guide the PMF analyses to more streamlined directions to ensure comparability between results obtained from different sites/instruments. According to this protocol, rolling PMF is performed following the latest and most advanced statistical features present within the Source Finder Professional (SoFi Pro) package (Datalystica Ltd., Villigen, Switzerland, (Canonaco et al., 2021, 2013)) integrated into the Igor Pro software (WaveMetrics Inc., Lake Oswego, OR, USA). Although subjective judgements cannot be avoided entirely, the developed protocol aims to minimise the number of decisions to be made by the user.

This highly time-resolved information of OA sources in Europe could substantially improve the development, validation, and prediction of regional/global air quality/climate models by providing extra independent information. These results could also be helpful to health-related studies when trying to accurately predict the toxicity of atmospheric aerosol since OA has significantly different health impacts depending on its origin (Daellenbach et al., 2020). Multiple years of data are finally needed to assess the impact of particulate matter sources on morbidity and mortality due to chronic exposure (Liu et al., 2019; Yang et al., 2019). Eventually, this work will provide valuable information for policymakers to take the most effective mitigation measures for aerosol-related environmental problems. Overall, this study presents a comprehensive overview of OA sources across Europe by following a thoroughly-designed and harmonised protocol (Section 2.3). Specifically, the seasonal/spatial variability of OA sources regarding time series and source profiles are unfolded in the following sections.

## 2  Measurements and Instrumentation

This study is the main outcome of the Chemical On-Line cOmpoSition and Source Apportionment of fine aerosoL (COLOSSAL) project (https://www.costcolossal.eu/), based on measurements performed within ACTRIS. In total, 22 year-long datasets were used here from 14 different countries via ACSM/AMS since 2013 (*Fig. 1*). This study includes data from 18 Q-ACSM (quadrupole ACSM, (Ng et al., 2011b)), 3 ToF-ACSM (Time-of-Flight ACSM (Fröhlich et al., 2013)), and 1 C-ToF-AMS (compact time-of-flight AMS (Drewnick et al., 2005)).

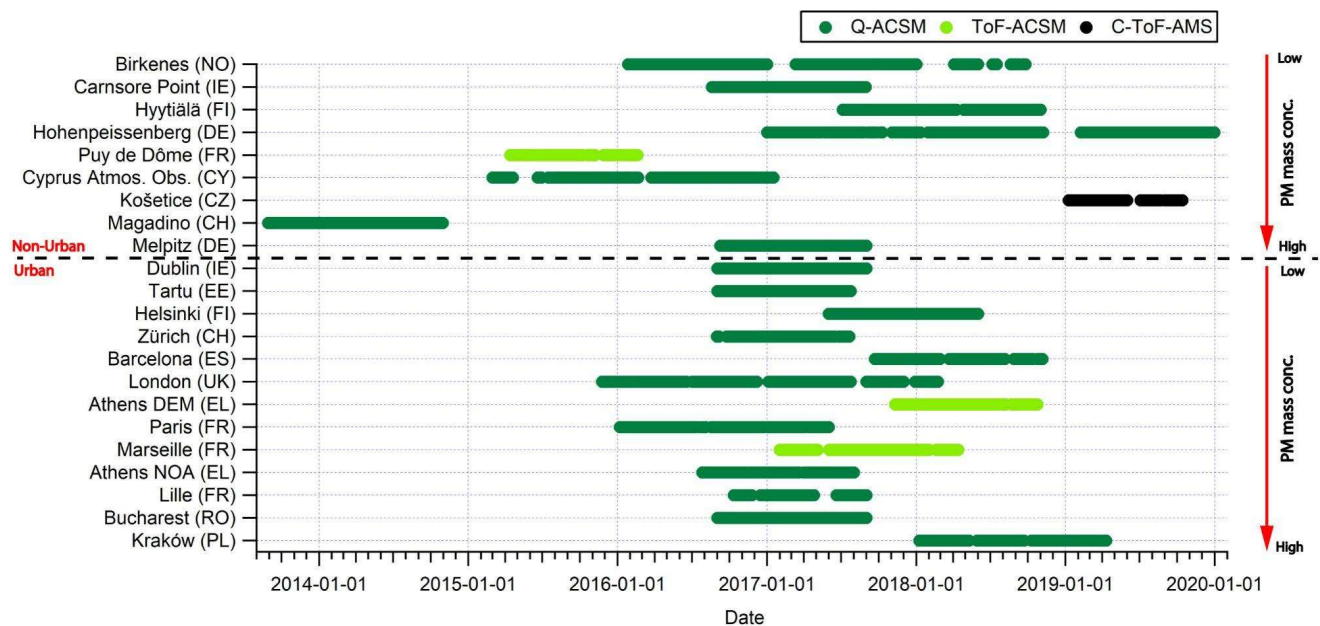

**Fig. 1.** ACSM/AMS measurement periods considered in this study.

Overall, the ACSM and AMS considered here apply similar techniques. Briefly, the air is passing through a critical orifice into an aerodynamic lens, where atmospheric aerosol is focused and accelerated (the smaller the aerodynamic size, the higher the velocity) into a vacuum chamber (10-5 Torr) and impacts on the surface of a standard vaporiser heated at 600 °C. The resulting vapours are then ionised by electron impact (electric ionisation at 70 eV), and these ions are further extracted into the detector to be characterised using the mass spectrometer. Compared to the AMS, the ACSM is more robust, affordable, and easier to operate, making it suitable for long-term monitoring purposes. However, it cannot measure the size-resolved chemical composition and its time and mass resolutions are poorer compared to the AMS. Most of the ACSMs deployed here participated in an inter-comparison activity conducted by the Aerosol Chemical Monitor Calibration Centre (ACMCC) at SIRTA (https://sirta.ipsl.fr/) and reported consistent results as long as proper calibrations were conducted (Crenn et al., 2015; Freney et al., 2019). One of the objectives of the COLOSSAL project is to deliver a harmonised standard operating procedure (SOP) for ACSM (COLOSSAL, 2021), and most of the 22 datasets were collected by following this SOP. However, the recommended relative ionisation efficiency (RIE) calibration procedures have varied over the long-time span of these datasets (2013-2019). For instance, some of the datasets conducted RIE calibration only on specific *m/z* values (jump scan) as recommended earlier, instead of scanning the entire mass range (10 to 150 amu) of the mass spectrometer (Freney et al., 2019). Considering the nitrate interference on the $CO_2^+$ signal at *m/z* 44 (so-called Pieber effect) is time-dependent (Freney et al., 2019; Fröhlich et al., 2015; Pieber et al., 2016) and *m/z* 44 is not measured in jump scan RIE calibrations, it is thus impossible to do a post-correction consistently. Therefore, none of the datasets was corrected for the Pieber effect. However, it is assumed that such artefact would not represent more than 25% (at a higher maximum, e.g., during severe ammonium nitrate pollution episodes) of total OA readings by AMS/ACSM used for the present study.

Complementary to ACSM/AMS measurements, equivalent black carbon (eBC) was also monitored at all sites using filter-based absorption photometers. It was typically measured using Multi Angle Absorption Photometer (MAAP) (Thermo) or Aethalometer Model AE33 (Magee Scientific) devices and predominantly with default settings as proposed by the manufacturer (i.e., with no extra data correction procedures). In the case of multi-wavelength instruments (e.g., AE33 or former AE31 devices), eBC concentrations were reported from measurements at 880 nm, and solid/liquid fuel-burning eBC subfractions (noted $eBC_{wb}$ and $eBC_{ff}$, respectively, hereafter) were distinguished from each other based on the application of the so-called Aethalometer model (Sandradewi et al., 2008; Zotter et al., 2017).

The 22 sampling sites are classified based on their geographic locations as urban (13 sites, including four flagged as suburban: Athens DEM, Lille, Paris, and Bucharest) or non-urban (9 sites, Table S1). The chemical composition of some of these 22 datasets have already been reported (Barreira et al., 2021; Bressi et al., 2021). This study focuses on the overview of OA source apportionment results and includes new sites in our analysis. More details about source apportionment results at some of these sites can be found in the following published papers: (Barreira et al., 2021; Canonaco et al., 2021; Chazeau et al., 2022, 2021; Chen et al., 2021; Farah et al., 2021; Heikkinen et al., 2021; Lin et al., 2019; Minguillón et al., 2015; Petit et al., 2021; Poulain et al., 2020; Stavroulas et al., 2019; Tobler et al., 2021; Via et al., 2021; Yttri et al., 2021; Zhang et al., 2019).

## 2.1 Positive matrix factorisation (PMF) and multilinear engine (ME-2)

PMF has been customarily performed to conduct source apportionment of ambient aerosol data (e.g., ACSM/AMS data) in many previous studies (Lanz et al., 2007; Ulbrich et al., 2009; Zhang et al., 2011). PMF model was first introduced by (Paatero and Tapper, 1994) as follows:

$$x_{ij} = \sum_{k=1}^{p} g_{ik} \times f_{kj} + e_{ij} \quad (1)$$

where $x_{ij}$ is the elements of the matrices for the measurements, $g_{ik}$ is the factor time series, $f_{kj}$ is the factor profiles, and $e_{ij}$ is the PMF residuals. The subscripts *i, j*, and *k* represent time, *m/z*, and a discrete factor, respectively. The superscript, *p* represents the number of factors. PMF finds the model solution by using the least-squares algorithm by iteratively minimising the following quantity:

$$Q = \sum_{i=1}^{n} \sum_{j=1}^{m} \left(\frac{e_{ij}}{\sigma_{ij}}\right)^2 \qquad (2)$$

where $\sigma_{ij}$ is the measurement uncertainty.

However, the PMF model does not provide a unique solution, which is usually referred to as rotational ambiguity (Paatero et al., 2002). Specifically, the model can deliver the exact same quantity of Q for a combination of the matrices **G** (time series) and **F** (factor profiles) and for a combination of their rotations $\underline{G}$ and $\underline{F}$ (with $\underline{G} = G \cdot T$ and $\underline{F} = T^{-1} \cdot F$). In this case, Q is the same, despite the solution being possibly entirely different. Even though the non-negativity constraints limit the number of allowed rotations, there are still many possible rotations and thus solutions. In order to reduce the rotational ambiguity of the PMF model, Paatero and Hopke (2009) proposed a multilinear engine (ME-2) algorithm, which allows the addition of *a priori* information into the model (e.g., source profiles or time series of external data) to prevent the model from unrealistic rotations and to generate more unique solutions. Using a priori information allows the user to guide the model towards an environmentally reasonable solution. Canonaco et al. (2013) implemented an ME-2 solver (Paatero, 1999) into the Igor-based software package, Source Finder (SoFi). SoFi enables the users to have enhanced rotational control over the factor solutions by imposing constraints via, e.g., the *a*-value approach on one or more elements of **F** and/or **G** (Paatero and Hopke, 2009). For instance, the *a* value (ranging from 0 to 1) is the tolerated relative deviation of a factor profile ($f_k$) or time series ($g_k$) from the chosen a priori input profile ($F_k$) or time series ($G_k$) during the iterative least-square minimization, as demonstrated in Equations 3a and 3b (Canonaco et al., 2013):

$$f_k = F_k \pm a \cdot F_k \qquad (3a)$$

$$g_k = G_k \pm a \cdot G_k \qquad (3b)$$

## 2.2 Rolling PMF analysis with bootstrap and random a-value approach

Conventional PMF is conducted over the whole dataset, with the assumption that the OA source profiles are static, which can lead to high errors when it comes to long-term datasets considering that OA chemical fingerprints are expected to vary over time (Paatero et al., 2014). For instance, (Canonaco et al., 2015) showed a substantial seasonal variability of oxygenated organic aerosol (OOA) factor profiles. Parworth et al. (2015) first proposed to run PMF analysis on a smaller time window (e.g., 14 days) to roll over the whole dataset with a certain step. This technique was further refined by Canonaco et al. (2021). The rolling PMF window mechanism allows the PMF model to adapt the temporal variations of the source profiles (e.g., biogenic versus biomass burning influences on OOA), which usually provides well-separated OA factors. In addition, with the help of the bootstrap strategy (Efron, 1979) and the random *a*-value approach, users can estimate the statistical and rotational uncertainties of the PMF results (Canonaco et al., 2021).

## 2.3 A standardised protocol of SA for a long-term dataset using SoFi Pro

This work presents a standardised protocol to identify main OA components. These guidelines work well for all 22 datasets despite the various pollution sources and OA levels at the different stations. All 22 datasets were analysed by following the protocol described in this study to minimise users' subjectivity. *Figure 2* provides the general working flow of this protocol, while the more detailed step-by-step guideline is summarised in *Table 1*. Detailed explanations of each step are unfolded in the following subsections.

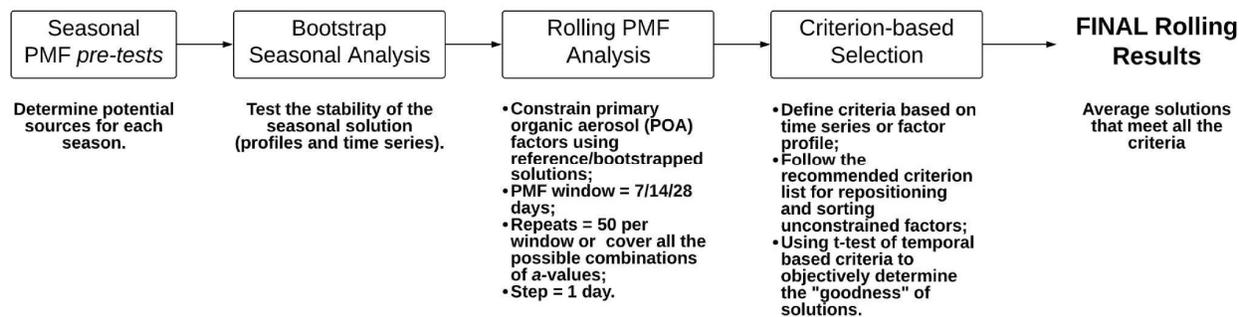

**Fig. 2.** Flow diagram of the standardised procedure for rolling PMF.

**Table 1.** Step-by-step protocol for running rolling PMF.

| | Detailed steps |
|---|---|
| **1. Seasonal PMF *pre-tests*** | 1.1. Unconstrained runs (2-8 factors) |
| | 1.2. Test for the presence of OA factors (in order of hydrocarbon-like OA (HOA), biomass burning OA (BBOA), cooking-like OA (COA), coal combustion OA (CCOA), special local factors (if applicable), and OOA factors) |
| | 1.3. Residual analysis (no structural patterns in diel profiles/time series/mass spectra) |
| | 1.4. *a*-value sensitivity analysis for constrained factors (i.e., primary OA factors (POAs)) |
| **2. Seasonal bootstrap analysis** | 2.1. Constrain the POAs and site-specific factor(s) in mass spectra retrieved from base case (the satisfactory solutions from seasonal *pre-tests*) |
| | 2.2. Combine the bootstrap resampling strategy with random *a*-values with an upper *a*-value of 0.4-0.5 for POAs and site-specific factor(s) and a repeat of 100-1000 times |
| | 2.3. Use the same technique as mentioned in step 4 (see later) to filter out PMF runs that are not environmentally reasonable |
| | 2.4. Quality check for the bootstrapped solution |
| **3. Rolling PMF** | 3.1. Constrain primary factor MS using published profiles or averaged site-specific profiles from seasonal bootstrapped solutions |
| | 3.2. Constrain site-specific factor MS (if those exist) using random *a*-value approach within a range of 0-0.4 and a step of 0.1. An upper limit of 0.5 for BBOA *a*-value could be considered |
| | 3.3. Enable bootstrap and set the length of the PMF window (7, 14 or 28 days) |
| | 3.4. Set the range of the number of factors based on the number of factors obtained during the seasonal analysis |
| **4. Criterion-based selection for PMF runs (Table S2)** | 4.1. Define a sorting criterion for more oxidised-oxygenated OA (MO-OOA) ($f_{44}$ for the MO-OOA) in case there are two unconstrained OOA factors |
| | 4.2. Selection criteria of PMF solutions based on correlations with external tracers |
| | 4.3. Selection criteria of PMF solutions based on time series (e.g., hours of COA, explained variation of key ions by specific factor) |
| | 4.4. Define the "best" PMF runs using the relevant/appropriate statistical tests (e.g., *t*-test approach) |
| | 4.5. Optimize time windows (compare non-modelled points and $Q/Q_{exp}$ among different time windows) |

### 2.3.1 Seasonal PMF *pre-tests*

To effectively implement rolling PMF analysis, knowing potential sources for each season for any given site is crucial. Seasonal PMF *pre-tests* allow us to retrieve reasonable seasonal PMF results (so-called base case). The first step of the *pre-tests* is to conduct unconstrained PMF with varying factors from 2 to 8 for each season. Based on the time series, diel patterns, factor profiles, and correlation with external data, the number of factors for each season could potentially be pre-determined. Examining a solution up to a relatively high number of factors (8 in this case) is crucial as some primary factors (e.g., HOA, BBOA, COA, and CCOA) might only appear in the PMF solution when the number of factors increases. Given the similar chemical fingerprints of some of the primary sources, it is possible that for a low number of factors, they remain mixed (e.g., HOA and COA). In this section, the approach used to identify each source type effectively is covered in the following paragraphs.

For a site potentially impacted by traffic, it is suggested to constrain the HOA mass spectrum from Crippa et al. (2013) using a tight *a* value (0.05-0.1) with a narrower range of factors (3-7). Typically, the HOA factor has a pronounced diel pattern with distinct morning and evening rush hour peaks, and the HOA factor profile is typically similar among different sites (Crippa et al., 2014). However, when the diel pattern does not show the typical variability expected from traffic emissions, one should consider using the HOA mass spectrum from unconstrained PMF runs or loosen the *a value* for HOA from Crippa et al. (2013). When a BBOA-like factor exists in both unconstrained and HOA-constrained runs, the "local" BBOA spectrum retrieved from these runs are recommended to be used as a constraint/reference profile in the next step of the PMF analysis. This is because of the relatively large spatial variabilities of BBOA factor profiles (Crippa et al., 2014). The expected BBOA factor usually has a pronounced contribution of *m/z* 29, 60, and 73 signals and a distinct diel pattern with high concentration during nighttime. If the BBOA factor was not present in previous steps, it should be checked if $f_{60}$ (i.e., the fraction of *m/z* 60 to the total organic mass) is above the background level of 0.3% (Cubison et al., 2011), also having a clear temporal pattern beyond the noise. However, it should not be the only criterion to determine the existence of BBOA (the background level of $f_{60}$ is instrument-dependent). Constrained PMF runs with a reference BBOA spectrum (Crippa et al., 2013; Ng et al., 2011a) and a relatively high *a* value (0.3-0.5) need to be performed to seek more proof of its existence by (i) comparing the solution without a BBOA factor; (ii) by checking the correlation factor between HOA vs eBC$_{ff}$, BBOA vs eBC$_{wb}$; and (HOA+BBOA) vs eBC$_{total}$; (iii) by checking if the solution has smaller scaled residuals of *m/z* 60, etc. One should keep in mind that the model performance metrics should determine the choice of reference BBOA profile.

If different slopes in $f_{55}$ vs $f_{57}$ plots at different hours of the day point towards the presence of COA (Mohr et al., 2012), it should be constrained tightly using the corresponding spectrum from Crippa et al. (2013) with an *a* value of 0.05-0.1. Then, it should be checked if the diel pattern (mass concentration or mass fraction) is reasonable, i.e., it peaks during the time of expected cooking-related activities (noon/afternoon and evening peaks). The COA factor is typically only present in urban environments close to residential and commercial areas.

For an environment with potential coal combustion sources, looking for a "local" CCOA factor mass spectrum from unconstrained or HOA-constrained PMF runs is recommended. However, it is typically challenging to identify the CCOA factor with ACSM data when its contribution is not significant because of the relatively low *m/z* resolution and since the mass range (up to 100/120 for Q-ACSM) does not include polyaromatic hydrocarbons (PAHs). In addition, the similar spectrum pattern of hydrocarbon ions (e.g., $C_nH_{2n-1}^+$ and $C_nH_{2n+1}^+$) between HOA, BBOA and CCOA makes it challenging for PMF to resolve these two factors (Sun et al., 2016). Therefore, the existence of CCOA should be justified at least by the most significant contribution of *m/z* 115 (mainly $C_9H_7^+$) and the absence of the morning rush hour peak.

Sometimes, PMF can also help picking up a site-specific factor with special fingerprints in previous steps (besides some common POAs and OOAs). In that case, the key ions should be checked to investigate potential sources, then the fragmentation table to understand how these key ions are calculated. Most importantly, OOA components should not be constrained as they are expected to vary season by season due to different precursor sources and oxidation capacity of the atmosphere. Typically, more and less oxidised-oxygenated OA (MO-OOA and LO-OOA, respectively) are found within the unconstrained factor(s). Here, the MO-OOA and low volatility oxygenated OA (LV-OOA) are interchangeable. The same applies to LO-OOA and semi volatile-oxygenated OA (SV-OOA). But we refer to these two OOA factors as MO-OOA and LO-OOA in the rest of the text. However, a static number of OOA factors for different seasons is highly recommended if it is environmentally reasonable since it remains challenging to objectively justify the exact transition period for the different number of factors at different periods.

Scaled residuals should be monitored throughout the PMF analysis. The scaled residuals' daily cycle structure might indicate missing or badly separated sources. In addition, spikes and structural patterns in the time series of the scaled residuals always require extra attention, as they suggest some high uncertainties of the current model for these points/periods or instrument issues. Also, if the constrained profiles cause systematic patterns in the residuals, they should be reconsidered or tested with a larger *a* value. Last but not least, the *a*-value sensitivity analysis for constrained factors should be conducted to optimise (i.e., towards enhanced correlation with externals, reasonable factor profiles, and small scaled residual, etc.) the constrained factor to the dataset of interest. Then, once the reasonable PMF solution has been determined (so-called base case result), the mass spectra for all constrained factors are input factors for bootstrap analysis in the next step.

### 2.3.2 Seasonal bootstrap analysis

A bootstrap resampling strategy (Efron, 1979) is recommended to test the stability of the base case solutions. Therefore, all the mass spectra of constrained factors (i.e., POAs and site-specific factors) should be constrained using the random $a$-value technique (Canonaco et al., 2021; Chen et al., 2021) with an upper $a$ value of 0.4-0.5 and repeats of 100-1000. Next, the same technique mentioned in Section 2.3.4 should be used to filter out "incorrect" solutions (not environmentally reasonable). As a next step, the quality of the averaged solution of selected PMF runs should be checked in terms of the uncertainties of factor profiles, time series, and percentage of selected runs. If the bootstrap solution shows significant uncertainties, the base case needs to be re-evaluated. The mass spectra for all constrained factors resulting from the bootstrap phase should be saved and used for the rolling PMF analysis in the next step.

### 2.3.3 Rolling PMF

The mass spectra from bootstrapped solutions are recommended as the reference profiles to constrain POA and site-specific factors during rolling PMF. BBOA is known as the most spatiotemporally variable factor compared to all other POA factors (Crippa et al., 2014). Also, considering that the highest mass concentrations of BBOA occur in winter, the BBOA mass spectrum from the bootstrapped winter solution is recommended as the constraint, as it is more representative of the dataset. Alternatively, the published profiles (HOA and COA from Crippa et al. (2013) and BBOA from Ng et al. (2011a) or Crippa et al. (2013)) could be used (Canonaco et al., 2021). Canonaco et al. (2021) suggested using a random $a$-value technique (randomly select $a$ values for each constraint within a certain range) and bootstrap resampling strategy to estimate the rotational uncertainties of rolling PMF. Based on the seasonal bootstrap analysis, the upper $a$ value for the site-specific factors can be determined by the seasonal variation and uncertainties. Canonaco et al. (2021) showed that an upper value of 0.4 for POAs is sufficient to cover the temporal variabilities. However, an upper $a$ value of 0.5 for BBOA is suggested when high temporal variabilities of the BBOA mass spectrum are expected. When the number of factors is not identical for all the seasons, the rolling PMF should be conducted with both n and n+1 factors over the entire dataset.

### 2.3.4 Criteria-based selection for PMF runs

With the large number of PMF runs expected for rolling analysis (e.g., >15,000 runs for a one-year dataset with 30-min time-resolution), inspecting each single PMF run is not feasible. Therefore, a criterion-based selection should be used to (i) evaluate the quality of the PMF runs quantitatively and relatively objectively and (ii) sort out the unconstrained factors in the same order for further averaging. The criteria-based selection has been explicitly explained in Canonaco et al. (2021). In short, SoFi Pro enables the user to define criteria based on the time series and/or factor profiles to select environmentally reasonable solutions. In addition, this criterion-based selection function can also serve to reposition unconstrained factors as unconstrained factors can appear in random order in a different iteration of the PMF. The inexact sorting criteria can result in a mixing of the unconstrained factors. Therefore, it is crucial to use the most representative sorting criterion, i.e., $f_{44}$ for the MO-OOA (criterion #7 in Table S2), as suggested by Chen et al. (2021). At the same time, it is also recommended to monitor $f_{44}$ in MO-OOA and $f_{43}$ in LO-OOA to reject zero-values of these two criteria. The criterion of $f_{44}/f_{43}$ for MO-OOA is not recommended because it could accept PMF solutions with smaller $f_{44}$ in MO-OOA than LO-OOA when $f_{43}$ is extremely small. Statistical tests such as the *t*-test (Chen et al., 2021) for time series based criteria (#4, #5 and #7 in Table S2) and correlation-based criteria (#1, #2 and #3 in Table S2) should be performed to minimise subjective decisions. With a *p* value $\leq$ 0.05, it is possible to select PMF runs with statistically significantly higher scores compared to the same criterion for other factors (for time series based criteria) or statistically higher correlation with each other (for correlation-based criteria). This technique allows the user to select environmentally reasonable PMF solutions with minimum subjective judgements. Last but not least, the optimum length of the time window should be determined by minimising non-modelled data points and $Q/Q_{exp}$ while applying the same criteria and thresholds to these PMF runs with different time windows (7, 14 or 28 days) (Canonaco et al., 2021; Chen et al., 2021). Based on our study, a 14-day window size is the most commonly selected one, which is consistent with previous studies (Canonaco et al., 2021; Chen et al., 2021).

### 2.3.5 Special cases and limitations of the current protocol

For all the 22 ACSM/AMS datasets, this standardised protocol works well in general. However, different numbers of factors at different periods remain challenging for rolling PMF. There are three special cases that this protocol could not cover. Specifically, the BBOA factor is not present in the warm period for the Barcelona, Cyprus, and Marseille datasets. However, this protocol could not cover such situations with a proper criterion to objectively include/exclude certain OA factors (e.g., BBOA) in certain time periods. The distribution of the correlation between BBOA and $eBC_{wb}$ was utilised for the Marseille data, which appears to have a bimodal Fisher distribution. Thus, the 10[th] percentile results from the separated distributions were used as thresholds to define the existence of BBOA (Chazeau et al., 2022). For the Barcelona and Cyprus dataset, the criterion to decide the existence of BBOA is the explained variation of $f_{60}$ by BBOA. A *t*-test was conducted with the null hypothesis that the variation of $f_{60}$ explained by BBOA is not significantly larger than that of other factors. The presence of BBOA is only considered when the *p* value was ≤ 0.05. In addition, as discussed before, a different number of factors often suffer from more uncertainties at the edge of the transition period by averaging different numbers of factor solutions simultaneously. One strategy to avoid averaging over different numbers of factor solutions is to unselect any data point in a range of *edge ± window length/2*. However, it could potentially lead to relatively more non-modelled points during the transition period. Therefore, keeping a static number of factors in the rolling analysis as much as possible is recommended if that is environmentally feasible. Thus, it remains challenging to objectively define the transition point to an improved source apportionment for rolling PMF analysis with a different number of OA factors.

## 3 Results and Discussions

*Figure 3* provides an overview of the mean OA mass fractions and their main components at 22 stations across Europe. Overall, the total $PM_1$ (sum of OA, eBC, nitrate ($NO_3$), sulfate ($SO_4$), ammonium ($NH_4$), and chloride (Chl)) mass concentration has an average of 9.7 ± 7.9 µg/m³, with generally higher concentrations at urban stations (brown circles, avg = 12.2 ± 9.3 µg/m³) compared to non-urban ones (green circles, avg = 6.2 ± 3.3 µg/m³). Kraków is the most polluted site (40.4 µg/m³), and Birkenes is the cleanest (1.3 µg/m³). The OA contribution to the total submicron aerosol ranges from 21 to 75%, which is consistent with previous results based on shorter campaigns (Jimenez et al., 2009). For other main chemical species, eBC, $NO_3$, $SO_4$, $NH_4$, and Chl exhibit an average contribution to the total $PM_1$ of 10.0%, 15.0%, 16.2%, 9.9% and 1.2%, respectively. eBC and $NO_3$ show higher contributions at the urban sites (12.1% and 16.2%) than at the non-urban ones (6.8% and 13.3%).

## 3.1 Overview of the primary and secondary OA

Primary and secondary OA factors are identified for each station in *Fig. 3*. Overall, the well-known POA factors have been resolved, including HOA and BBOA. All datasets identify HOA except Hyytiälä (non-urban). BBOA is resolved in 19 datasets (12 urban and 7 non-urban sites) except for Hyytiälä, Puy de Dôme, and Helsinki. In general, both HOA and BBOA present considerable components of the total submicron aerosol with average contributions of 5.0% and 5.6%, respectively. Also, they both show higher contributions at the urban sites (HOA: 5.7% and BBOA: 6.3%) than at the non-urban ones (HOA: 4.0% and BBOA 4.5%). COA is identified in three southern European cities (i.e., Athens, Marseille, and Barcelona), a megacity (London), and a central European city (Zürich). It has an average contribution to the total $PM_1$ of 6.3%. CCOA is resolved in Kraków (Tobler et al., 2021) and Melpitz (non-urban) with contributions of 5.8% and 6.9%, respectively. SFOA, which likely originates from peat and coal combustion, appears at the two Irish sites (Dublin (urban) and Carnsore Point (non-urban)), with contributions of 12.2% and 6.0%, respectively. In addition, local factors (particular of the monitoring site) are highlighted in this study: an *m/z* 58-related OA (58-OA) in Magadino (1.2%); a coffee roaster OA factor in Helsinki (3.0%); a sea salt factor at Carnsore Point (1.5%); cigarette smoking OA (CSOA) in Zürich (Qi et al., 2019; Stefenelli et al., 2019); and a mixed ship-industry factor in Marseille (2.2%, Chazeau et al., 2022). Finally, two secondary OOA factors (MO-OOA and LO-OOA) are present at all sites except for Birkenes (MO-OOA only). The MO-OOA and LO-OOA contribute to the total $PM_1$ with an average of 22.0% and 13.1%, respectively. Both MO-OOA and LO-OOA show drastic differences among urban (16.7% and 10.6%) and non-urban sites (29.6% and 17.2%), which is expected since more primary sources are present in the urban environment. When summing up MO-OOA and LO-OOA (Total OOA), **Fig. 3** suggests that secondary OA is the main contributor to total submicron PM (average = 34.5%, range from 11.7 to 62.4%) and dominates OA (average = 71.1%, range from 47.3 to 100%) across Europe.

In addition, the resolved OA factors have been validated using available external data (**Table S3**). Regarding the PMF errors (Equation (6) in Canonaco et al. (2021)), they are estimated by logarithmic probability density functions (pdf) of the standard deviations of each time point i divided by the mean concentration of each time point i for corresponding OA factors. The PMF errors of major OA components are presented in **Table S4**. In general, POAs often have smaller PMF errors than OOA factors since they are always constrained.

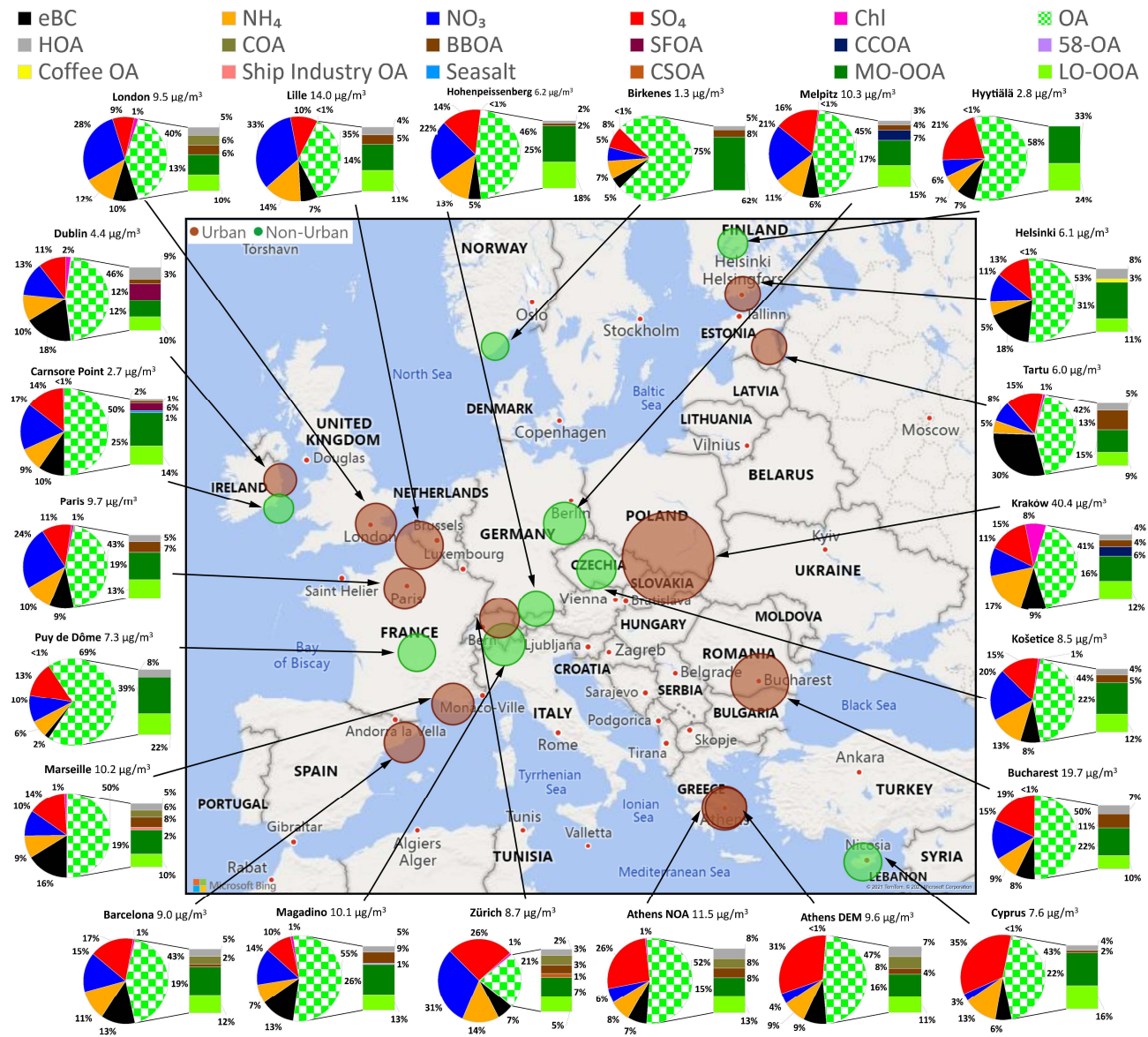

**Fig. 3.** Submicron particulate matter ($PM_1$) mass concentration (in µg/m³) and mass fractions of non-refractory inorganic species, equivalent black carbon (measured by online filter-based methods), and organic aerosol measured with the 22 ACSM/AMS at multiple locations in Europe covering all seasons. The size of the markers corresponds to the $PM_1$ mass concentration. The brown colour of the marker indicates an urban site, while the green marker indicates a non-urban site. The light green/white texture of the pie charts is the organic aerosol (OA) fraction in $PM_1$, and the bar charts represent the contributions of each OA factor to the total OA mass (The map was generated from Microsoft Power BI).

## 3.2 OA composition changes as a function of OA loading

To understand how the OA composition changes under different loadings, each dataset is divided into ten bins containing the same number of points based on the OA mass concentrations. As shown in *Fig. 4*, apart from Zürich and Helsinki, all urban and non-urban sites (Magadino and Melpitz) report larger POA contributions under high OA loadings. It is most likely because primary emissions substantially contribute to OA mass concentrations in relatively polluted areas under stagnant conditions (Tobler et al., 2021; Zhang et al., 2019). Specifically, HOA shows a relatively constant contribution even when the OA mass concentration increases at non-urban sites, while it increases as the OA mass concentration increases at urban sites. This suggests that traffic emissions still play a considerable role during high OA mass concentration periods, and it remains important even when the total OA mass concentration is low. In particular, all factors related to the combustion of solid fuel (i.e., BBOA, SFOA, and CCOA) show the most pronounced enhancement when OA mass loading increases, especially during winter seasons (*Fig. S2* in the Supplement), which suggests solid fuel combustion is the main driver for the polluted episodes. The sea salt factor at Carnsore Point, in turn, has the highest contribution in the bottom 10 per cent of OA mass loadings since high wind speeds favour high sea salt emissions and low OA mass concentrations at the same time.

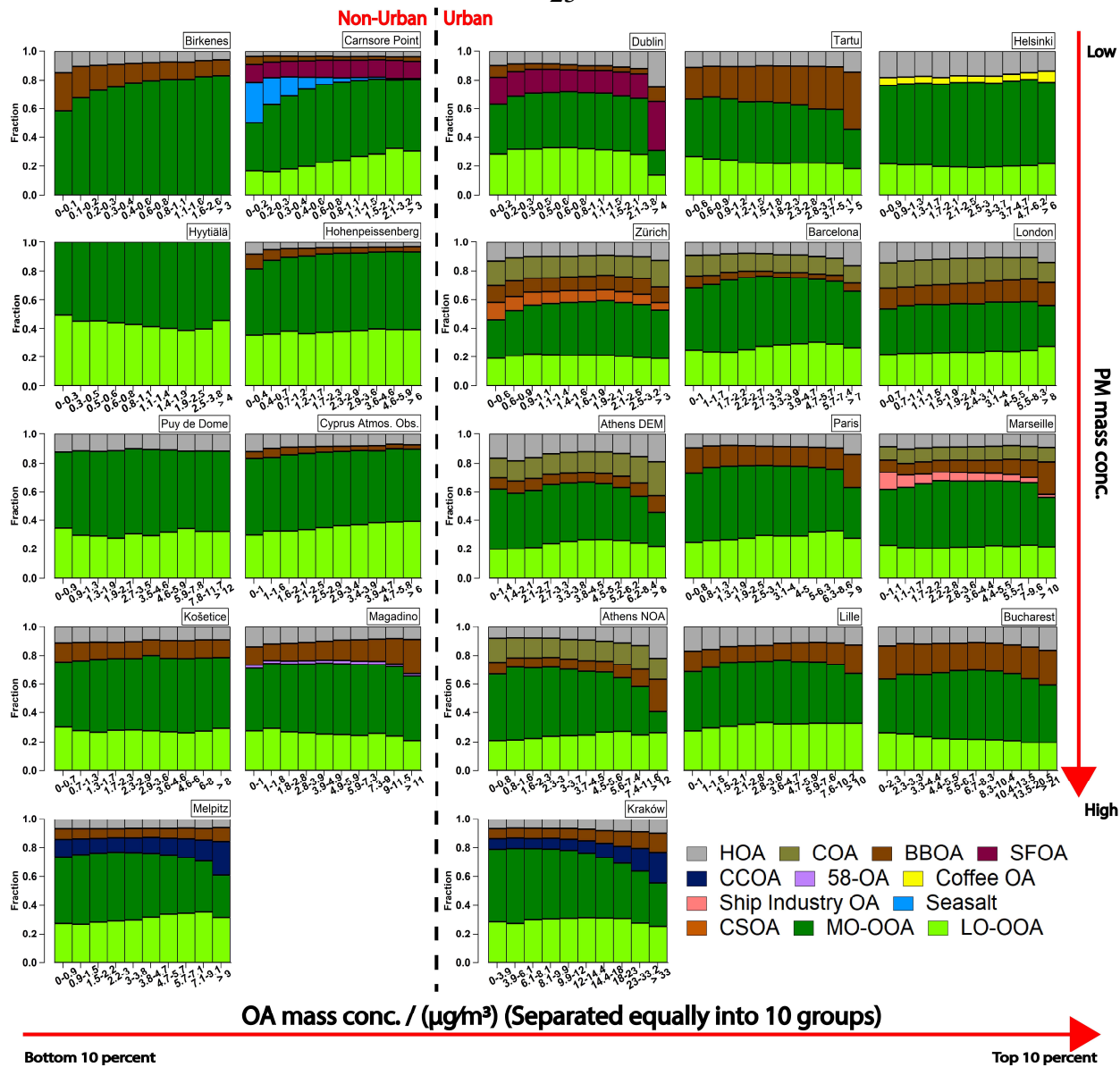

**Fig. 4** Mass fraction of each organic aerosol (OA) component for 22 stations with 10 equally distributed bins (based on OA mass concentration). Non-urban and urban sites have been separated by the dashed vertical line divided into left and right panels. The sites are sorted based on the submicron total particulate matter ($PM_1$) mass concentration. In general, oxygenated OA (OOA, dark and light green bars) is dominant at all sites over the whole mass range. However, with an increase of the total OA mass concentration, an increase of primary OA is often observed.

## 3.3 Diel cycles for resolved OA components

The highly time-resolved long-term ACSM/AMS data allow investigating the diel cycles of the OA components. ***Figure 5*** shows the averages (solid lines) and standard deviations (shaded areas) of diel profiles of the major OA components (HOA, BBOA, COA-like, MO-OOA, LO-OOA and Total OOA), normalised by the annual average of the corresponding OA component mass concentration for each site at both non-urban and urban sites. Birkenes and Hyytiälä datasets are not included because they only have 4-hour and 3-hour time resolutions, respectively.

HOA shows a distinct pattern at urban sites with characteristic morning and evening rush-hour peaks. By contrast, at non-urban locations, HOA does not follow the same pattern, indicating that this factor is likely associated with transported traffic emissions or with non-traffic primary hydrocarbon emissions at these sites. COA, as mentioned resolved at six urban sites, shows distinct noon and evening peaks with minor standard deviations, which suggests small spatial variabilities of cooking emissions. BBOA has a similar diel cycle at urban and non-urban sites with reduced values during the day and a marked evening peak, which indicates that most likely, (residential) heating emissions are the main contributor to BBOA. However, at the urban sites, the evening peak of BBOA is more pronounced than at the non-urban sites, which suggests that the urban BBOA is more local and synchronised to domestic heating than the non-urban sites. The delayed morning peaks and broader span of non-urban BBOA diel profiles further indicate relatively distant BBOA sources at non-urban sites.

The MO-OOA diel trend at both urban and non-urban sites shows the most stable pattern. A slight decrease is observed during the night starting after 23:00 and continuing until 06:00–07:00, probably due to the decrease in the formation rates of MO-OOA in the absence of photochemical activity. By contrast, the MO-OOA concentrations increase slightly from morning to afternoon at the non-urban sites, potentially due to photochemistry. Moreover, the long-range transported origin could also play a role in this stable diel trend of MO-OOA. The diel cycles of LO-OOA at the urban and non-urban sites reveal nighttime maxima with a slight decrease at noon, suggesting local production or enhanced vapour partitioning onto pre-existing aerosol in the shallow nocturnal boundary layer. Urban LO-OOA shows much stronger evening peaks than non-urban sites, potentially caused by nighttime chemistry yielding urban OOA from POA oxidation. Kodros et al. (2020) and Tiitta et al. (2016) demonstrated how dark ageing of BBOA could potentially yield substantial amounts of OOA. Such ageing is likely to explain some of our observations at the urban sites (Zhang et al., 2020). Both urban and non-urban sites show small spatial variability for the Total OOA, but the diel cycle for urban sites has a more substantial evening peak again due to the strong LO-OOA increase. In general, the POA factors show more temporal variability than the OOA factors at both urban and non-urban sites.

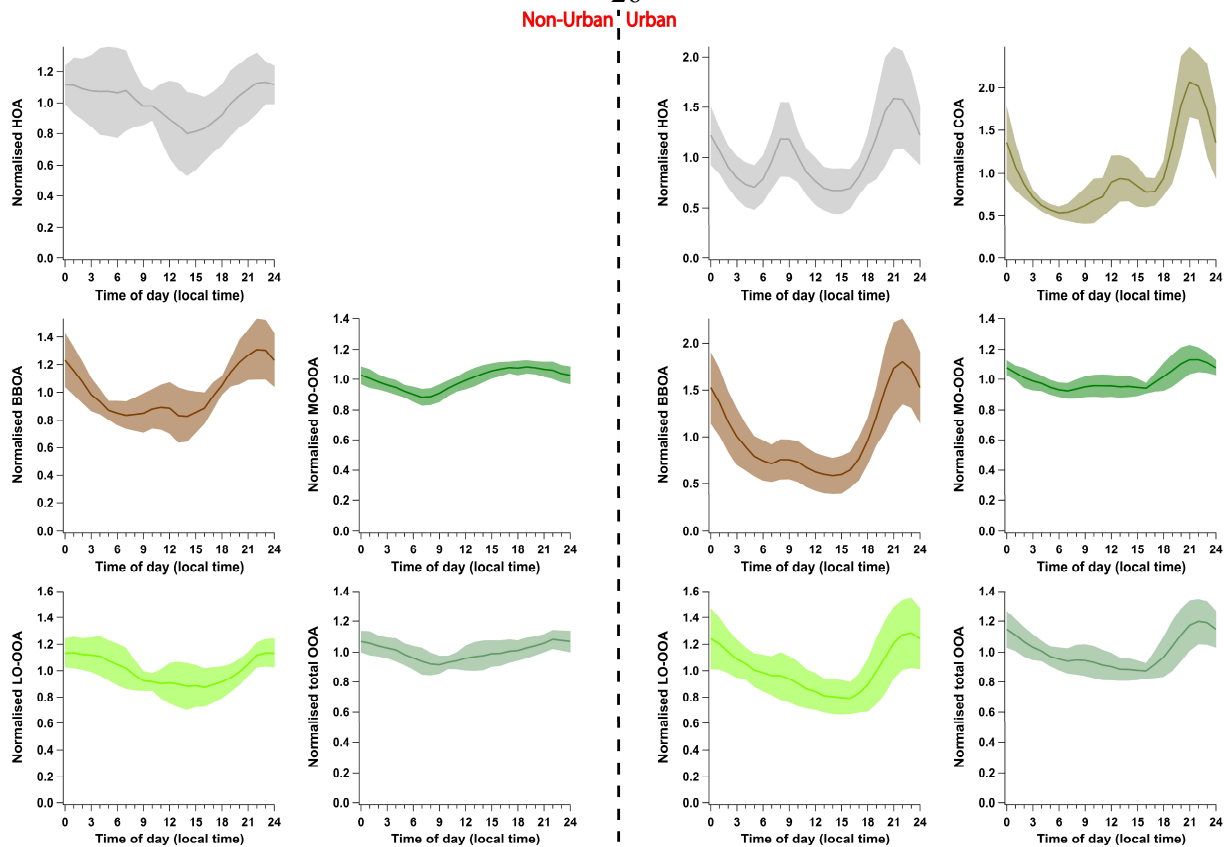

**Fig. 5.** Normalised diel profiles for HOA, COA, BBOA, MO-OOA, LO-OOA, and Total OOA (MO-OOA+LO-OOA). The solid lines and shaded areas represent the averages and standard deviations, respectively, of the diel profiles for the non-urban and urban datasets (normalised by the annual average of the corresponding OA component mass concentration for each site). The POA factors generally show more pronounced variability than the OOA factors. In addition, the urban sites have stronger patterns with less spatial variability than the non-urban sites.

In addition, *Fig. S4* shows the normalised weekly cycles for the non-urban and urban sites separately (with Birkenes and Hyytiälä datasets included among the non-urban sites). Compared with the diel cycles, the weekly ones are much weaker in general. In addition, POA shows a stronger variability compared to the OOA factors, similar to the averaged diel cycles (*Fig. 5*). HOA shows decreased values during the weekend than the weekdays at urban (-2.6%) and non-urban (-3.5%) sites. Except for HOA, the non-urban weekly cycles are much less pronounced than the ones at the urban sites. Specifically, BBOA increases during the weekend with 19.9% and 12.3% higher than the weekdays for urban and non-urban sites, respectively. This is because more wood-burning activities (e.g., open fire grills and residential heating) are expected during weekends (Fuller et al., 2014). COA shows a similar trend as it increases during the weekends (+18.8% at urban sites), suggesting that cooking activities are more pronounced. All OOA factors (MO-OOA, LO-OOA, and Total OOA) do not present strong weekly cycles (<6.3% difference between the weekday and weekend), with relatively larger spatial variability for OOA at the urban sites, indicated by larger standard deviations.

### 3.4 Spatial and seasonal variability of OA contributions

The time series of daily-averaged OA fractions for each site is shown in **Fig. S1**, which presents a big picture of the entire source apportionment result. It indicates a significant spatial and temporal variabilities of OA contributions across Europe. To study the seasonal variation of OA and its sources, data was divided into four seasons: winter (DJF: December, January, and February), spring (MAM: March, April, and May), summer (JJA: June, July, and August), and autumn (SON: September, October, and November). ***Figure 6*** indicates a relatively small spatial variability of the relative OA contributions at both urban and non-urban datasets, and there is no clear pattern between the OA fraction and PM$_1$ loading. Urban sites have higher POA contributions in OA than non-urban sites. However, each dataset shows an apparent seasonal variability with higher POA contributions and mass concentrations in cold seasons than in warm ones (***Fig. 6*** and ***Fig. S5***). The contributions of POA also appear to be higher when the total PM$_1$ mass concentration increases at all non-urban sites, as shown in ***Fig. 6***.

Specifically, urban sites show higher HOA contributions (overall average contribution of 12.7 ± 2.9%) than non-urban sites (7.4 ± 2.7%), which is expected due to more traffic emissions in urban areas. Moreover, for both urban and non-urban sites, the HOA contribution shows a distinct seasonality with the lowest contribution in summer (8.8 ± 3.4%) and the highest in winter (12.0 ± 4.8%), and similar contributions in spring (10.9 ± 4.3%) and autumn (11.5 ± 4.6%). This might be due to the lower boundary layer with stagnant conditions in the cold season favouring the accumulation of primary and local pollutants and/or reduced photochemistry. In addition, the heating related sources (i.e., BBOA, CCOA, and SFOA) are obviously more pronounced during the cold seasons than the warm ones. Specifically, BBOA has an average contribution of 8.3 ± 4.7% in summer but 16.9 ± 8.4% in winter. CCOA was only found in Kraków and Melpitz, and its contributions varied from season to season with substantially enhanced contribution during winter (Kraków: 18.2% and Melpitz: 23.1%) compared to summer (Kraków: 4.5% and Melpitz: 8.7%). The drastic seasonal variations in Kraków are due to the widespread use of coal-burning for residential heating purposes in winter. In Melpitz, the coal combustion is less dependent on local sources, but most likely, emissions from Poland or other eastern European countries are rapidly transported by advection in winter, leading to the observed seasonality. For the Dublin dataset, the SFOA (heavily affected by both peat and coal combustion sources (Lin et al., 2019, 2018)) shows an enhanced contribution in winter (32.9%) and a decreased contribution in summer (13.2%). The SFOA in Carnsore Point station shows less seasonality because of the absence of local sources but still has a higher contribution in winter (13.3%) than in summer (9.2%). The COA contribution shows almost no seasonality ranging from 14.3 ± 2.7% in summer to 15.4 ± 3.3% in autumn at the six urban sites where this factor is resolved, suggesting that the cooking emission contribution is constant and non-negligible in European cities in all seasons. The specific contributions of all OA components for all datasets and all seasons are summarised in Table S5, in which all POA except HOA, BBOA and COA are summed up as "other OA".

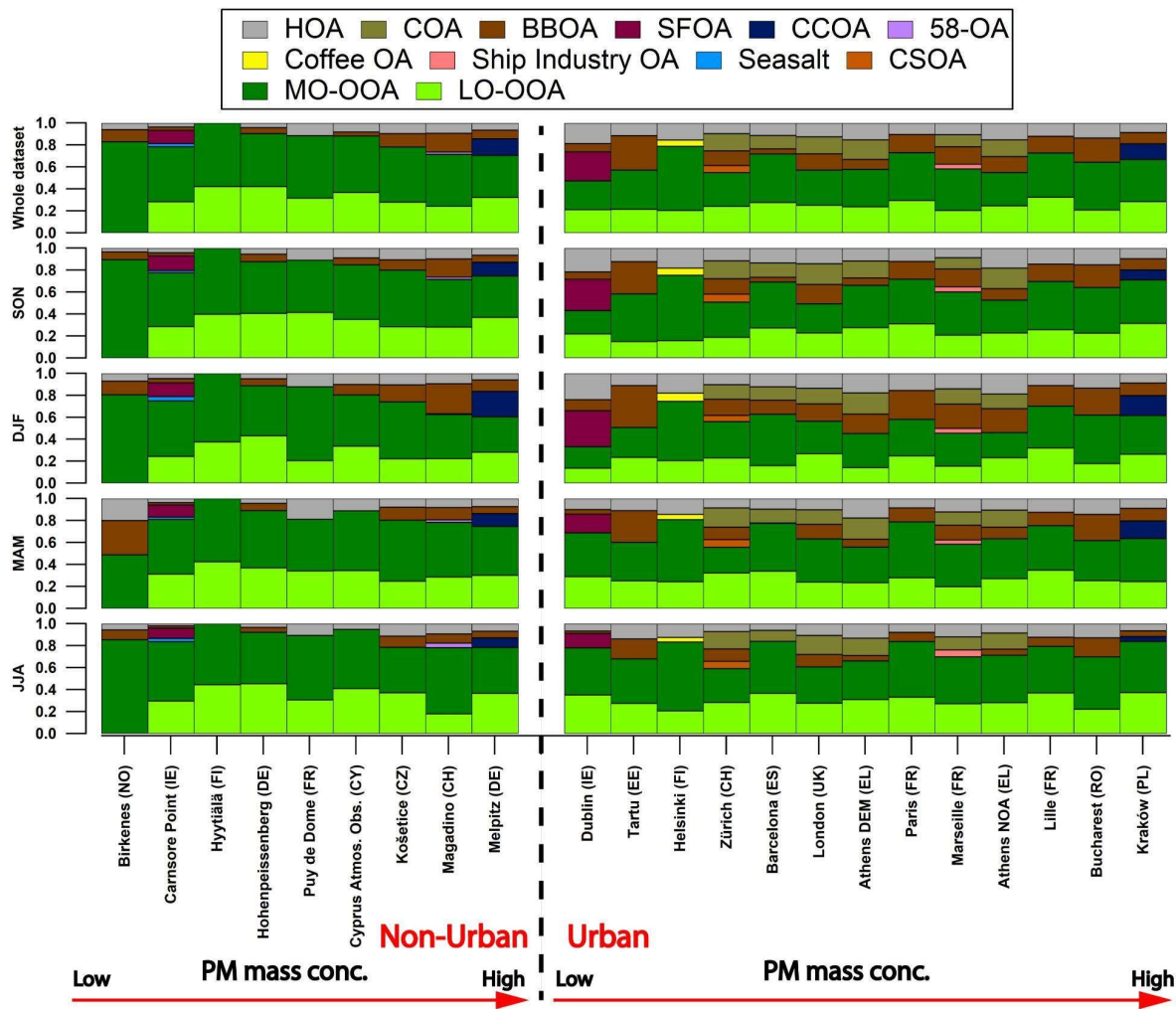

**Fig. 6.** Relative contributions of all OA components at each station grouped by season (see text). The stations are categorised into non-urban (left) and urban sites (right), where each subset is ordered by $PM_1$ mass concentration.

### 3.5 Spatial and seasonal variabilities of $f_{44}$ vs $f_{43}$ in the OOA factors

Canonaco et al. (2015) analysed the seasonal variability of major ion intensities (i.e., *m/z* 44 and *m/z* 43) in the OOA factors in Zürich. They suggested that biomass burning emissions, the most important precursor of SOA in that city during winter, cause the LO-OOA factors to be at the left half of Sally's triangle (Ng et al., 2010) within the $f_{44}$ vs $f_{43}$ space, as shown for biomass burning emissions by Heringa et al. (2011). On the contrary, biogenic SOA from terpene oxidation may "push" the LO-OOA factors to the right side of the triangle in summer (Pfaffenberger et al., 2013).

This study explores the seasonality of $f_{44}$ vs $f_{43}$ for MO-OOA, LO-OOA and the Total OOA factors for both seasons at all sites, as shown in **Fig. 7.** It shows that the rolling PMF provides a good separation between MO-OOA and LO-OOA in both winter and summer (**Fig. 7a**), which is consistent with what Canonaco et al. (2021) and Chen et al. (2021) reported for two Swiss datasets. However, the positions of MO-OOA for the different stations in the $f_{44}$ vs $f_{43}$ space show large spatial variability (**Fig. 7a**), mainly attributed to the complex and various ageing processes in different locations under different meteorological conditions. In general, all LO-OOAs shift to the right side, though by a different extent during the summer (JJA) compared to the winter (DJF) months. This is more apparent when summing up the LO-OOA and MO-OOA into the Total OOA factors (**Fig. 7b**). The shift of LO-OOA and the Total OOA factors is most likely due to the enhanced biogenic emissions with higher temperatures during summer seasons (Canonaco et al., 2015).

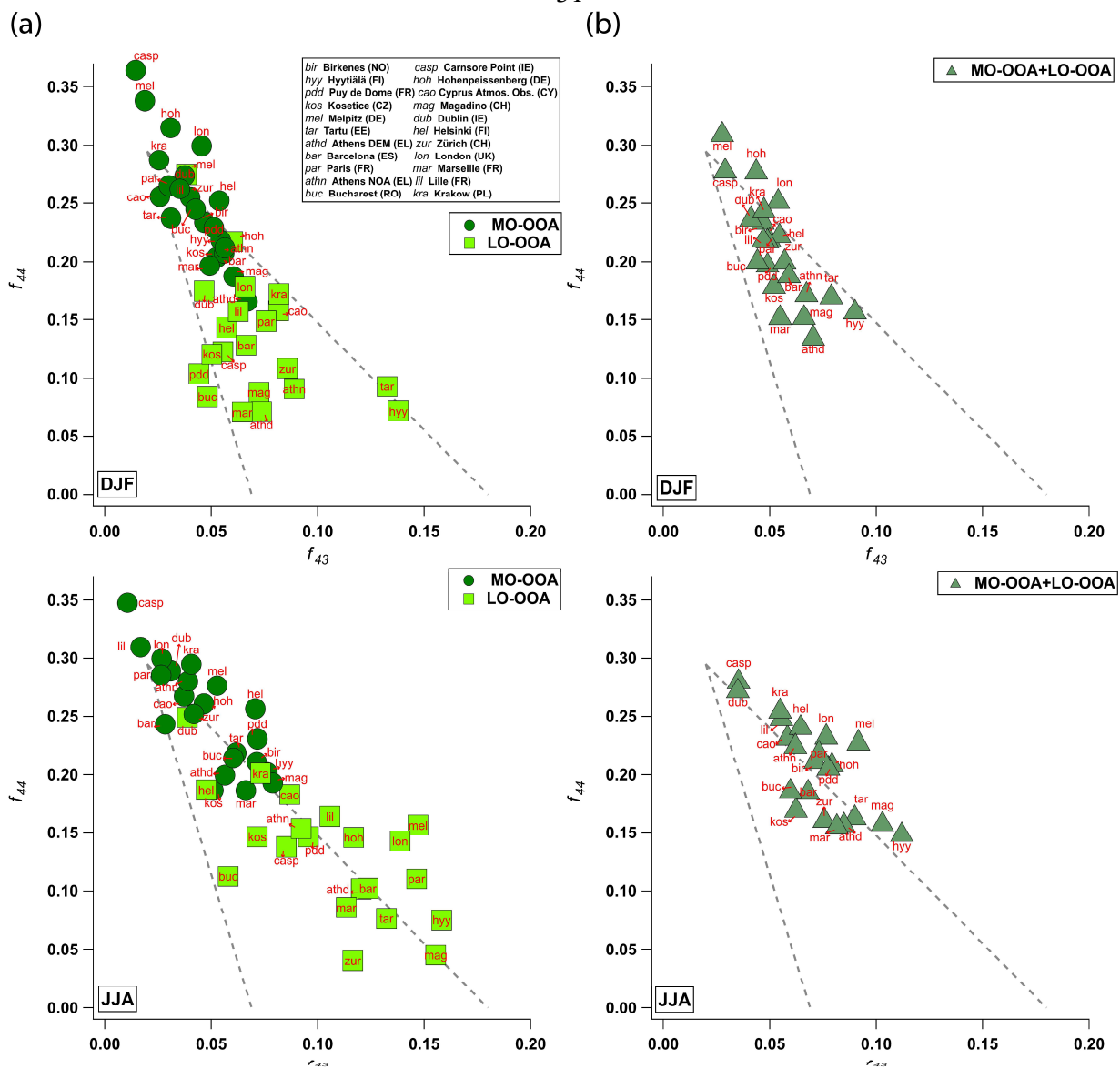

**Fig. 7.** (a) $f_{44}$ vs $f_{43}$ intensity in winter (top) and summer (bottom) for (a) resolved MO-OOA (dark green) and LO-OOA (light green); (b) the Total OOA (MO-OOA+LO-OOA).

At the same time, some site-specific ions could significantly influence the LO-OOA position. For instance, due to the pronounced biomass burning influence in Bucharest during summer, the LO-OOA factor has considerable contributions of *m/z* 55, 57, 60 and 73, which is accompanied by a low $f_{43}$ in LO-OOA during summer. The LO-OOA factors of the Tartu and Hyytiälä datasets stay on the right side of the triangle even during the winter season. Tartu appears to have a significant

BBOA contribution throughout the year (18.2% in summer). Considering biomass burning oxidises rapidly, the biomass burning influence is more pronounced in the MO-OOA factor during the whole year. That is why the LO-OOA factor in winter remains on the right side of the triangle. Hyytiälä is located in the boreal forest with significant biogenic SOA formation in summer (Heikkinen et al., 2021; Yli-Juuti et al., 2021), which explains the high $f_{43}$ in summer. In addition, no POA factors were deconvolved from this dataset following the presented protocol. By utilizing machine learning techniques, Heikkinen et al. (2021) resolved a slightly aged POA factor that could neither be further separated into HOA nor BBOA. This factor appeared only in winter and coincided with a LO-OOA drop to near-zero loadings when utilizing k-mean clustering approach (Heikkinen et al., 2021). Therefore, it is very likely that the Hyytiälä LO-OOA shown in this study is influenced by aged POA in winter, which keeps the LO-OOA $f_{43}$ high. Also, the potential terpene emission from Korkeakoski sawmills (ca. 7 km NE of the monitoring station) could also keep the LO-OOA on the right side (Äijälä et al., 2017). The combined effects of high summertime biogenic SOA contribution to LO-OOA and wintertime POA mixing to LO-OOA could explain why LO-OOA in Hyytiälä always stays on the right side of the triangle. When considering the rest of the sites, both the LO-OOA and Total OOA factors generally shift to the right side of the triangle during summer compared to winter.

### 3.6 Spatial and temporal variabilities of key ions in resolved OA factors

This study also investigates key ions' spatial and monthly variabilities in common OA factors (i.e., *m/z* 55 and *m/z* 57 for HOA and COA; *m/z* 60 and *m/z* 73 for BBOA; *m/z* 44 and *m/z* 43 for MO-OOA, LO-OOA, and the Total OOA). All monthly intensities of these key ions in these OA factors were averaged across the 22 datasets to see possible monthly trends (***Fig. 8***). Overall, the key ions for HOA, COA, MO-OOA and Total OOA factors barely show a monthly trend. In contrast, $f_{60}$ and $f_{73}$ in BBOA are significantly higher in the cold months compared to the warmer months. The main reason is likely that levoglucosan and thus *m/z* 60 is not stable in the warm season as well as the change in the biomass burning source during different seasons (e.g., residential heating and outdoor open fire) (Bertrand et al., 2018; Bougiatioti et al., 2014; Xie et al., 2014). The most dominating ions (*m/z* 44 and *m/z* 43) in the LO-OOA factor show a relatively strong monthly trend compared with MO-OOA and Total OOA. Specifically, $f_{44}$ is smaller, and $f_{43}$ is higher in LO-OOA during the warm months, which is a further indication that the enhancement of biogenically-formed SOA could increase the intensity of $f_{43}$ in LO-OOA when the temperature is increased (which eventually "pushes" the LO-OOA factors to the right side of the triangle as presented in **Fig. 7**, (Canonaco et al., 2015)). Monthly trends of these key ions of these OA factors for each station are shown in ***Fig S6.***

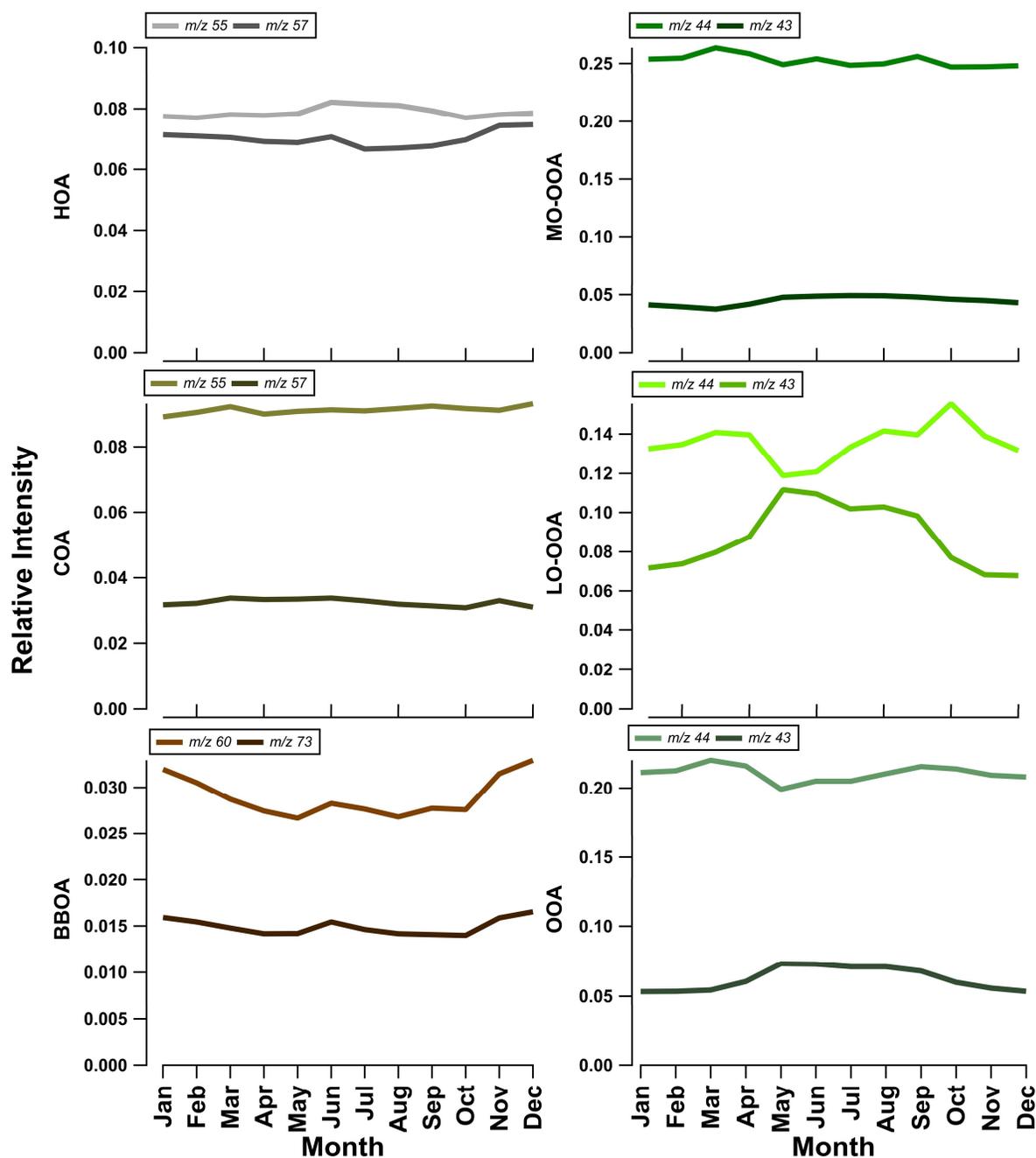

**Fig. 8.** Monthly average relative intensities of key ions in the corresponding factor profiles over all the datasets.

In order to compare the spatial variations of key ions in the different OA factors, the interquartile range (IQR) of the monthly site-averaged intensities has been normalised by their median (***Table 2***). In general, $f_{55}$ for HOA and COA shows a relatively small IQR/median ratio with averages of 0.15 ± 0.03 and 0.16 ± 0.07, respectively. The $f_{57}$ for HOA and COA shows similar consistency across 22 sites with averages of 0.20 ± 0.03 and 0.09 ± 0.05, respectively. Overall, the HOA and COA factors are generally consistent across different locations. It agrees well with the previous findings reported by Crippa et al. (2014). However, the most essential fingerprint ions of BBOA, $f_{60}$ and $f_{73}$, appear to have the largest spatial variability among the POA factors with IQR/median ratio in a range of 0.28-0.39 and 0.20-0.36, respectively. This is expected since the type of wood and burning conditions, as well as chemical ageing, can affect the BBOA mass spectrum significantly (Bertrand et al., 2018; Bougiatioti et al., 2014; Grieshop et al., 2009; Heringa et al., 2011; Weimer et al., 2008; Xie et al., 2014). Thus, as discussed in a previous section, retrieving site-specific BBOA factor profiles using unconstrained PMF analysis instead of published BBOA factor profiles is strongly recommended.

**Table 2.** Normalised spatial variations for key ions using the interquartile range (IQR) divided by the medians of average monthly intensities across the 22 datasets.

| Factors | HOA | | COA | | BBOA | | MO-OOA | | LO-OOA | | Total OOA | |
|---|---|---|---|---|---|---|---|---|---|---|---|---|
| m/z | 55 | 57 | 55 | 57 | 60 | 73 | 44 | 43 | 44 | 43 | 44 | 43 |
| Jan | 0.13 | 0.21 | 0.11 | 0.04 | 0.36 | 0.29 | 0.30 | 0.46 | 0.52 | 0.46 | 0.34 | 0.34 |
| Feb | 0.16 | 0.19 | 0.12 | 0.04 | 0.33 | 0.35 | 0.25 | 0.57 | 0.60 | 0.54 | 0.31 | 0.29 |
| Mar | 0.15 | 0.19 | 0.10 | 0.11 | 0.37 | 0.25 | 0.28 | 0.45 | 0.57 | 0.39 | 0.26 | 0.39 |
| Apr | 0.12 | 0.22 | 0.06 | 0.10 | 0.36 | 0.25 | 0.27 | 0.69 | 0.60 | 0.60 | 0.32 | 0.39 |
| May | 0.17 | 0.24 | 0.09 | 0.11 | 0.36 | 0.23 | 0.34 | 0.83 | 0.86 | 0.53 | 0.37 | 0.36 |
| Jun | 0.19 | 0.17 | 0.11 | 0.17 | 0.34 | 0.20 | 0.27 | 0.81 | 0.63 | 0.53 | 0.40 | 0.33 |
| Jul | 0.19 | 0.21 | 0.18 | 0.17 | 0.32 | 0.21 | 0.27 | 0.65 | 0.38 | 0.40 | 0.29 | 0.29 |
| Aug | 0.21 | 0.27 | 0.27 | 0.14 | 0.28 | 0.22 | 0.26 | 0.69 | 0.50 | 0.43 | 0.29 | 0.30 |
| Sep | 0.19 | 0.21 | 0.24 | 0.04 | 0.32 | 0.23 | 0.25 | 0.70 | 0.94 | 0.41 | 0.36 | 0.33 |
| Oct | 0.11 | 0.18 | 0.25 | 0.06 | 0.39 | 0.21 | 0.27 | 0.49 | 0.46 | 0.59 | 0.30 | 0.26 |
| Nov | 0.11 | 0.13 | 0.19 | 0.08 | 0.34 | 0.36 | 0.19 | 0.60 | 0.50 | 0.55 | 0.21 | 0.41 |
| Dec | 0.12 | 0.22 | 0.14 | 0.02 | 0.33 | 0.36 | 0.16 | 0.77 | 0.74 | 0.42 | 0.29 | 0.39 |
| Mean | 0.15 ± 0.03 | 0.20 ± 0.03 | 0.16 ± 0.07 | 0.09 ± 0.05 | 0.34 ± 0.03 | 0.26 ± 0.06 | 0.26 ± 0.05 | 0.64 ± 0.13 | 0.61 ± 0.16 | 0.49 ± 0.08 | 0.40 ± 0.05 | 0.34 ± 0.05 |

The most dominating ion in MO-OOA, $f_{44}$, has a somewhat smaller IQR/median ratio, ranging from 0.16 to 0.34, and the $f_{43}$ in MO-OOA has an IQR/median ratio in a range of 0.45-0.83. This is because different datasets could have significantly variable ageing processes, precursors, and meteorological conditions, which appear to affect the degrees of oxygenation (the relative intensities of $m/z$ 44 and $m/z$ 43). In addition, MO-OOA remains dominated mainly by $f_{44}$ (average $f_{44}$ = 0.25), but with a much smaller $f_{43}$ intensity (average $f_{43}$ = 0.04). Therefore, the slight changes in the intensity of $f_{43}$ could have a larger effect on the IQR/median ratio for this ion.

Due to enhanced biogenic emissions, the $f_{43}$ in LO-OOA shows consistently larger intensities in the warmer months (May-September), as shown in **Fig. 8**. Consequently, the $f_{44}$ in LO-OOA decreases during the warm seasons and is related to the seasonal differences in $f_{44}$ vs $f_{43}$ of **Fig. 7.** In addition to the dynamic monthly trends observed in LO-OOA, there are also strong spatial variabilities for these two key ions (i.e., $m/z$ 44 and $m/z$ 43). The IQR/median ratios of $f_{44}$ and $f_{43}$ in LO-OOA are in the range 0.38-0.94 and 0.39-0.60, respectively. This is expected considering LO-OOA has never been constrained, and various ageing processes, precursors, and meteorological conditions could contribute to the large spatial variabilities. When we sum up the LO-OOA and MO-OOA to the Total OOA factor, $m/z$ 44 is still the dominating ion with an average of 0.21, but it is smaller than that of MO-OOA alone due to the much smaller $f_{44}$ in LO-OOA with an average of 0.14. Overall, $f_{43}$ in the Total OOA factor still shows an increasing trend during warm seasons like the LO-OOA factor, which indicates that the effect of enhanced biogenic emissions on the intensity of $f_{43}$ in the Total OOA might be rather considerable. Moreover, $f_{43}$ in the Total OOA shows relatively smaller spatial variabilities (compared to MO-OOA and LO-OOA) with an IQR/median ratio of 0.26-0.41. The $f_{44}$ in the Total OOA has slightly larger spatial variabilities than MO-OOA with an IQR/median ratio of 0.21-0.40, but it is still more stable than LO-OOA. It is expected considering the large spatial variabilities in LO-OOA. However, the sum of the OOA factors shows little monthly trends except for the increasing $f_{43}$ (thus, decreasing in $f_{44}$) in warmer months (May-September).

# 4 Conclusion

A state-of-the-art standardised protocol for source apportionment of long-term ACSM/AMS organic aerosol mass spectrum datasets was developed. Our protocol was validated systematically and strictly applied to 22 sites with year-long measurements. It demonstrates the consistency of this protocol with comprehensive source apportionment results, even though each dataset was analysed by each research group individually. Our source apportionment strategy has been significantly improved compared to conventional seasonal PMF by utilising rolling windows, bootstrap, and ME-2 techniques, which were first introduced by Canonaco et al. (2021). As addressed by Chen et al. (2021) and Tobler et al. (2021), this strategy allows us to retrieve robust source apportionment results by considering temporal variations of source profiles. Importantly, the success of the rolling mechanism is an essential step to make real-time source apportionment possible (Chen et al., 2022). However, the current protocol/strategy remains challenging to objectively define the transition point to an improved source apportionment for rolling PMF analysis when a different number of OA factors is necessary for different periods. More tools should be tested to address this challenge in the near future.

Overall, this work provides a comprehensive overview of the OA sources in Europe with highly time-resolved source apportionment results. The OA fraction in $PM_1$ is high (23-75%) across Europe and seasons. With the help of this advanced source apportionment strategy, many common POA factors have been resolved, including HOA, COA, BBOA, CCOA, and SFOA, together with secondary OOA factors, i.e., MO-OOA and LO-OOA in these 22 datasets. Moreover, some local OA components have been identified at specific stations, like a Coffee roastery OA factor in Helsinki, a Ship Industry OA factor in Marseille, a sea salt factor in Carnsore Point, and a cigarette smoke OA factor in Zürich. The OOA factors together constitute the main contributor (47.3-100%) to OA and generally show more stable diel and weekly cycles than POA factors. The contributions of POA increase with increasing total OA mass concentration in most of the polluted regions. It suggests that the control of primary emissions could help mitigate OA mass concentration or at least decrease the likelihood of highly polluted episodes. Also, most POA factors show enhanced contribution/mass concentrations during cold seasons compared to warm seasons due to residential heating. Lower boundary layer heights (lower temperature) combined with stagnant conditions can readily cause the accumulation of pollutants. In particular, HOA (traffic emissions) is a non-negligible OA source with a rather consistent contribution across different stations (10.7 ± 3.8%). Six urban sites display a significant COA factor (i.e., two Athens datasets, Zürich, London, Barcelona, and Marseille) with an overall average contribution to OA of 14.5 ± 2.5%. Moreover, most of the datasets show a resolved BBOA component (except Hyytiälä, Puy de Dôme, and Helsinki) with important contributions to OA (annual average:12.4 ± 6.9%), which increases significantly in winter with a contribution of 16.9 ± 8.4%. Melpitz and Kraków present a CCOA factor (annual average: 14.7 ± 0.8%, winter average: 20.6 ± 3.5%), while SFOA is found to be heavily affected by peat and coal combustion sources at Carnsore Point and Dublin (annual average: 19.3 ± 10.4%, winter average: 22.7 ± 14.4%). All of these results confirm that the reduction of solid fuel-burning for residential heating is one of the key leverages to mitigate fine PM levels in Europe, especially in winter.

This study also provides a comprehensive overview of the spatial and temporal variabilities of commonly resolved OA factors (i.e., HOA, COA, BBOA, MO-OOA, LO-OOA, and Total OOA). Overall, the MO-OOA, LO-OOA, and Total OOA factors vary significantly spatially and seasonally. This is expected since the ageing processes, abundances/types of precursors and meteorological conditions can differ both temporally and spatially. Regarding the seasonality of the LO-OOA factor, most of the datasets agree well with the findings reported by Canonaco et al. 2015), with increasing $f_{43}$ (thus, decreasing $f_{44}$) intensity in warm seasons likely due to the enhanced biogenic emissions.

Moreover, with the help of the rolling PMF technique, time-dependent OA factor profiles have been retrieved. Therefore, this study also investigates the monthly trends (averaged over 22 datasets) and corresponding variabilities (IQR/median ratio) across sites for key ions in the commonly resolved OA factors. While these key ions barely show monthly trends in HOA, COA, MO-OOA, and Total OOA, BBOA key ion ($f_{60}$ and $f_{73}$) intensities increase during the cold seasons due to the abundance of biomass burning sources and the lower reactivity of levoglucosan. The increased $f_{43}$ of LO-OOA during warm seasons is most likely due to enhanced biogenic SOA formation. In terms of spatial variabilities, key ions for HOA and COA factors show a small IQR/median ratio with a range of 0.06-0.27 due to both factor contributions being consistent (if present) as reported in previous studies. However, the key ions for the BBOA factor show a relatively larger spatial variability with the IQR/median ratio ranging from 0.20 to 0.39, which suggests potentially different combustion conditions, type of woods, etc., contribute to this variability in the real-world scenarios. In addition, the $f_{43}$ intensities in MO-OOA and LO-OOA show large spatial variability, with the IQR/median ratio ranging from 0.45 to 0.83 and 0.39 to 0.60, respectively. The $f_{44}$ in LO-OOA has a large IQR/median ratio range of 0.38-0.94, but the $f_{44}$ in MO-OOA is rather less varied across sites with a relatively small IQR/median ratio range of 0.16-0.34. This is expected since OOA factors are never constrained combined with the complexities of ageing processes in different locations and meteorological conditions.

With the help of this state-of-the-art source apportionment protocol, this study has retrieved highly time-resolved, long-term (>9 months), and robust OA source information consistently, with minimum subjective judgements. This highly time-resolved comprehensive OA source information can be useful inputs/constraints to improve/validate climate, health, and air quality models.

Finally, this work suggests that policymakers should address the major combustion sources (i.e., biomass burning, coal combustion, and peat) that affect air quality in Europe. Besides, more attention should be paid to the traffic source, even though it is quite constant across places because it is significant in organic aerosols and is also a proxy for non-exhaust emissions. Due to regional transport, decreasing the emissions of primary factors would also decrease the secondary factors observed at the non-urban sites.

# 5 Acknowledgement

This work would not have been possible without the following contributions: All co-authors were supported by the cost action of Chemical On-Line cOmpoSition and Source Apportionment of fine aerosol (COLOSSAL, CA16109). GC and AT were supported by a COST related project of the Swiss National Science Foundation, Source apportionment using long-term Aerosol Mass Spectrometry and Aethalometer Measurements (SAMSAM, IZCOZ0_177063), as well as the EU Horizon 2020 Framework Programme via the Research Infrastructures Services Reinforcing Air Quality Monitoring Capacities in European Urban & Industrial AreaS (RI-URBANS) project (GA-101036245), the ERA-PLANET projects SMURBS and iCUPE (grant agreement no. 689443). LP and SA were supported by ACTRIS (grant nos. ACTRIS (262254) and ACTRIS-2 (654109)). HC, VR, and JFB acknowledge financial support from the Labex CaPPA project, which is funded by the French National Research Agency (ANR) through the PIA (Programme d'Investissement d'Avenir) under contract ANR-11-LABX-0005-01, and the CLIMIBIO project, both financed by the Regional Council "Hauts-de-France" and the European Regional Development Fund (ERDF). In addition, The ATOLL (Lille), Longchamp (Marseille), and SIRTA (Paris region) platforms are part of the so-called LCSQA-CARA program, co-funded by the French Ministry of Environment. JV, CM, LM were supported by a grant of the Romanian Ministry of Research, Innovation and Digitalization, CNCS - UEFISCDI, project number PN-III-P1-1.1-TE-2019-0340, within PNCDI III and by Romanian National Core Program contract 18N/2019. European Union's Horizon 2020 research and innovation programme (project FORCeS under grant agreement No 821205). European Research Council (Consolidator grant INTERGRATE No 865799. European Research Council (and Starting grant COALA No 638703). RL, PP, JS and PV were supported by ICPF CAS acknowledges financial support from the MEYS's of the Czech Republic within the INTER-EXCELLENCE INTERCOST program under grant agreements No. LTC18068 and by under grant Actris CZ LM2018122, and by the GACR under grant P209/19/06110Y. IDAEA-CSIC acknowledges financial support from Generalitat de Catalunya (AGAUR 2017 SGR41), the Spanish Ministry of Science and Innovation through CAIAC project (PID2019-108990RB-I00) and FEDER funds, through EQC2018-004598-P. IDAEA-CSIC is a Centre of Excellence Severo Ochoa (Spanish Ministry of Science and Innovation, Project CEX2018-000794-S). The Cyprus Atmospheric Observatory team acknowledges funding from the European Union's Horizon 2020 Research and Innovation Programme (under grant agreement no. 856612) and the Cyprus Government. Operational program Competitiveness, Entrepreneurship and Innovation (RESEARCH-CREATE-INNOVATE) project code: T1EDK-03437. KRD acknowledges support by the SNSF Ambizione grant PZPGP2_201992. The LCE team acknowledges financial support from the PACA region (PRISM project; grant n°2017_08809). TROPOS has been supported by the German Federal Environment Ministry (BMU) grants F&E 370343200 (German title: "Erfassung der Zahl feiner und ultrafeiner Partikel in der Außenluft"),2008–2010, and F&E 71143232 (German title: "Trendanalysen gesundheitsgefährdender Fein- und Ultrafeinstaubfraktionen unter Nutzung der im German Ultrafine Aerosol Network (GUAN) ermittelten Immissionsdaten durch Fortführung und Interpretationder Messreihen"), 2012–2014. KS and ASk were supported by the subsidy from the Ministry of Science and Higher Education, grant numbers: 16.16.220.842 B02, 16.16.210.476 and by the EU Project POWR.03.02.00-00-I004/16. EF, JEP and OF gratefully acknowledge CNRS-INSU for supporting measurements performed at SIRTA and CO-PDD, within the long-term monitoring aerosol program SNO-CLAP, both of which are components of the ACTRIS French Research Instructure, and whose data is hosted at the AERIS data center (https://www.aeris-data.fr/). LS gratefully acknowledges DIM Qi² research network for financial support of her doctoral work. National University of Ireland Galway work was supported by the EPA Research Programme 2021-2030, AEROSOURCE Project (2016-CCRP-MS-31). The EPA Research Programme is a Government of Ireland initiative funded by the Department of Environment, Climate and Communications. GM acknowledges the support of the Slovenian Research Agency (grant no. P1- 0385). NOA team (IS, AB, NM (Nikolaos Mihalopoulos)) acknowledges support by the "PANhellenic infrastructure for Atmospheric Composition and climatE change" (MIS 5021516) which is implemented under the Action "Reinforcement of the Research and Innovation Infrastructure", funded by the Operational Programme "Competitiveness, Entrepreneurship and Innovation" (NSRF 2014-2020) and co-financed by Greece and the European Union (European Regional Development Fund). JS and MP acknowledge the support of the European Union's Horizon 2020 Research and Innovation Programme (under grant agreement no. 856612) and of the Cyprus Government.

Moreover, we also want to thank Krista Luoma for Hyytiälä aethalometer data. Helsinki Region Environmental Services Authority HSY and Metropolia University of Applied Sciences for providing auxiliary Helsinki site data. The authors gratefully acknowledge the Romanian National Air Quality Monitoring Network(NAQMN, www.calitateaer.ro) for NOx data providing.

# *Supplementary Information for:* European Aerosol Phenomenology - 8: Harmonised Source Apportionment of Organic Aerosol using 22 Year-long ACSM/AMS Datasets

Gang Chen[1], Francesco Canonaco[1,2], Anna Tobler[1,2], Wenche Aas[3], Andres Alastuey[4], James Allan[5,6], Samira Atabakhsh[7], Minna Aurela[8], Urs Baltensperger[1], Aikaterini Bougiatioti[9], Joel F. De Brito[10], Darius Ceburnis[11], Benjamin Chazeau[1,12,13], Hasna Chebaicheb[10,14], Kaspar R. Daellenbach[1], Mikael Ehn[15], Imad El Haddad[1], Konstantinos Eleftheriadis[16], Olivier Favez[14], Harald Flentje[17], Anna Font[18*], Kirsten Fossum[11], Evelyn Freney[14,20], Maria Gini[16], David C Green[18,19], Liine Heikkinen[15**], Hartmut Herrmann[7], Athina-Cerise Kalogridis[16], Hannes Keernik[21,22], Radek Lhotka[23,24], Chunshui Lin[11], Chris Lunder[3], Marek Maasikmets[21], Manousos I. Manousakas[1], Nicolas Marchand[12], Cristina Marin[25,26], Luminita Marmureanu[25], Nikolaos Mihalopoulos[9], Griša Močnik[27,28], Jaroslaw Nęcki[29], Colin O'Dowd[11], Jurgita Ovadnevaite[11], Thomas Peter[30], Jean-Eudes Petit[31], Michael Pikridas[32], Stephen Matthew Platt[3], Petra Pokorná[23], Laurent Poulain[7], Max Priestman[18], Véronique Riffault[10], Matteo Rinaldi[33], Kazimierz Różański[29], Jaroslav Schwarz[23], Jean Sciare[32], Leïla Simon[14,31], Alicja Skiba[29], Jay G. Slowik[1], Yulia Sosedova[2], Iasonas Stavroulas[9,32], Katarzyna Styszko[34], Erik Teinemaa[21], Hilkka Timonen[8], Anja Tremper[18,19], Jeni Vasilescu[25], Marta Via[4,35], Petr Vodička[23], Alfred Wiedensohler[7], Olga Zografou[16], María Cruz Minguillón[4***], and André S.H. Prévôt[1***]

[1]Laboratory of Atmospheric Chemistry, Paul Scherrer Institute, 5232 Villigen, Switzerland
[2]Datalystica Ltd., Park innovAARE, 5234 Villigen, Switzerland
[3]NILU - Norwegian Institute for Air Research, 2007 Kjeller, Norway
[4]Institute of Environmental Assessment and Water Research (IDAEA), Spanish Council for Scientific Research (CSIC), Barcelona, 08034, Spain
[5]Department of Earth and Environmental Sciences, University of Manchester, Manchester, UK
[6]National Centre for Atmospheric Science, University of Manchester, Manchester, UK
[7]Department of Chemistry of the Atmosphere Leibniz Institute for Tropospheric Research, Permoser Straße 15, 04318, Leipzig, Germany
[8]Atmospheric Composition Research, Finnish Meteorological Institute, P.O. Box 503, 00101, Helsinki, Finland
[9]Institute for Environmental Research and Sustainable Development, National Observatory of Athens, Palaia Penteli, 15236, Athens, Greece
[10]IMT Nord Europe, Institut Mines-Télécom, Univ. Lille, Centre for Energy and Environment, 59000 Lille, France
[11]School of Physics, Ryan Institute's Centre for Climate and Air Pollution Studies, National University of Ireland Galway, University Road, Galway, H91 CF50, Ireland
[12]Aix Marseille Univ., CNRS, LCE, Marseille, France
[13]AtmoSud, Regional Network for Air Quality Monitoring of Provence-Alpes-Côte-d'Azur, Marseille, France
[14]Institut National de l'Environnement Industriel et des Risques, Parc Technologique ALATA, 60550, Verneuil en Halatte, France
[15]Institute for Atmospheric and Earth System Research (INAR) / Physics, University of Helsinki, Helsinki, Finland

[16] Environmental Radioactivity Laboratory, Institute of Nuclear & Radiological Sciences & Technology, Energy & Safety, N.C.S.R. "Demokritos", 15310 Athens, Greece

[17] Deutscher Wetterdienst, Meteorologisches Observatorium Hohenpeißenberg, 82383 Hohenpeißenberg, Germany

[18] MRC Centre for Environment and Health, Environmental Research Group, Imperial College London, 86 Wood Lane, London, W12 0BZ, UK

[19] HPRU in Environmental Exposures and Health, Imperial College London, UK

[20] Laboratoire de Météorologie Physique, UMR6016, Université Clermont Auvergne-CNRS, Aubière, France

[21] Air Quality and Climate Department, Estonian Environmental Research Centre (EERC), Marja 4D, Tallinn, Estonia

[22] Department of Software Science, Tallinn University of Technology, 19086 Tallinn, Estonia

[23] Institute of Chemical Process Fundamentals of the CAS, Rozvojová 135/1, 16502 Prague, Czech Republic

[24] Institute for Environmental Studies, Faculty of Science, Charles University, Benátská 2, 12801 Prague, Czech Republic

[25] National Institute of Research and Development for Optoelectronics INOE 2000, 77125 Magurele, Romania

[26] Department of Physics, Politehnica University of Bucharest, Bucharest, Romania

[27] Condensed Matter Physics Department, J. Stefan Institute, Ljubljana, Slovenia

[28] Center for Atmospheric Research, University of Nova Gorica, Ajdovščina, Slovenia

[29] AGH University of Science and Technology, Faculty of Physics and Applied Computer Science, Department of Applied Nuclear Physics, Kraków, Poland

[30] Institute for Atmospheric and Climate Sciences, ETH Zürich, Zürich, 8092, Switzerland

[31] Laboratoire des Sciences du Climat et de l'Environnement, UMR 8212, CEA/Orme des Merisiers, 91191 Gif-sur-Yvette, France

[32] Climate & Atmosphere Research Centre (CARE-C), The Cyprus Institute, Nicosia, 2121, Cyprus

[33] Institute of Atmospheric Sciences and Climate (ISAC), National Research Council (CNR), 40129 Bologna, Italy

[34] AGH University of Science and Technology, Faculty of Energy and Fuels, Department of Coal Chemistry and Environmental Sciences, Kraków, Poland

[35] Department of Applied Physics, University of Barcelona, Barcelona, 08028, Spain

*Now at: IMT Nord Europe, Institut Mines-Télécom, Univ. Lille, Centre for Energy and Environment, 59000 Lille, France

**Now at: Department of Environmental Science & Bolin Centre for Climate Research, Stockholm University, Stockholm, Sweden

*** Correspondence to: María Cruz Minguillón (mariacruz.minguillon@idaea.csic.es) and André S. H. Prévôt (andre.prevot@psi.ch)

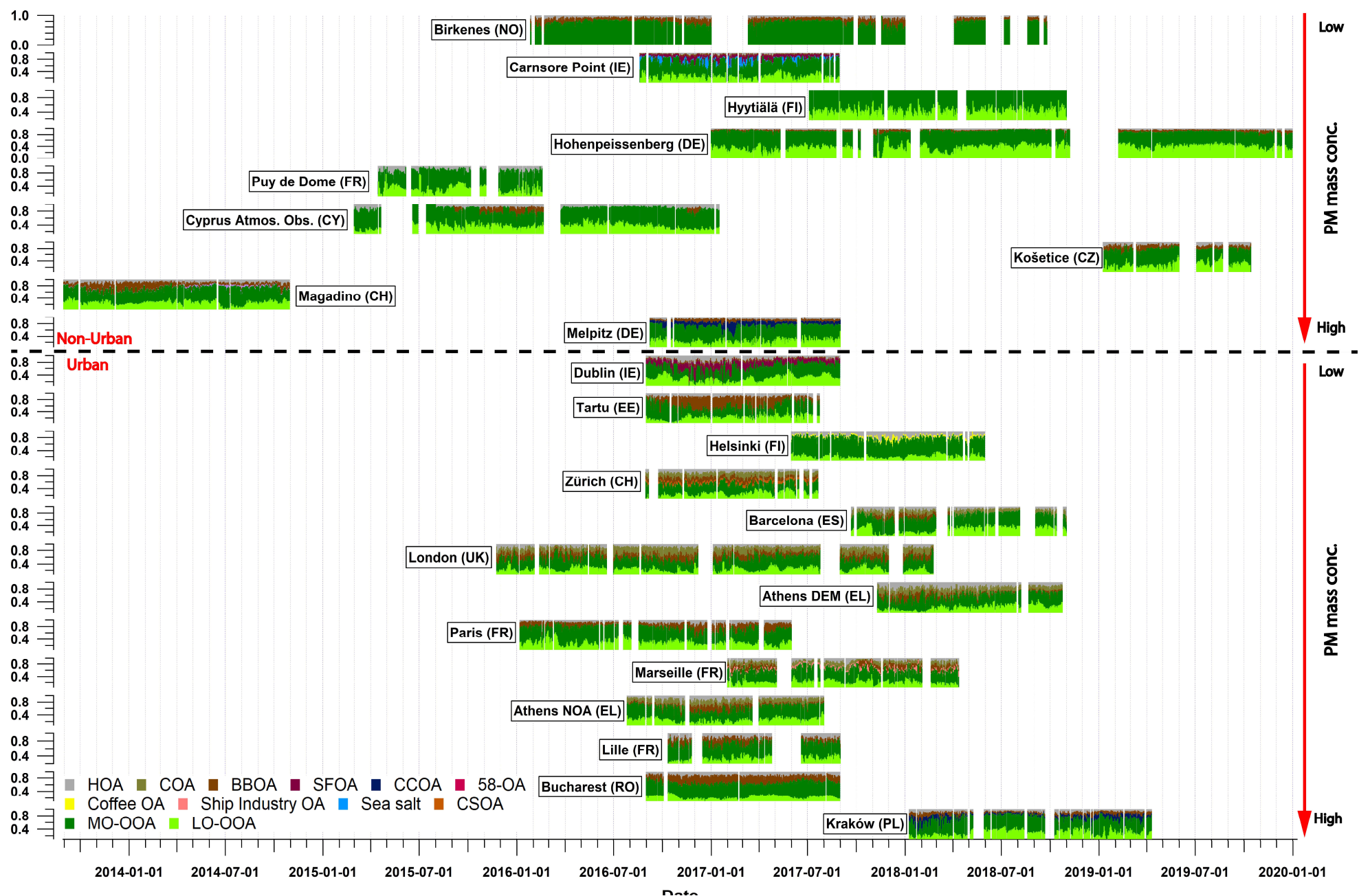

**Fig. S1.** Time series of the daily-averaged OA contribution for each factor.

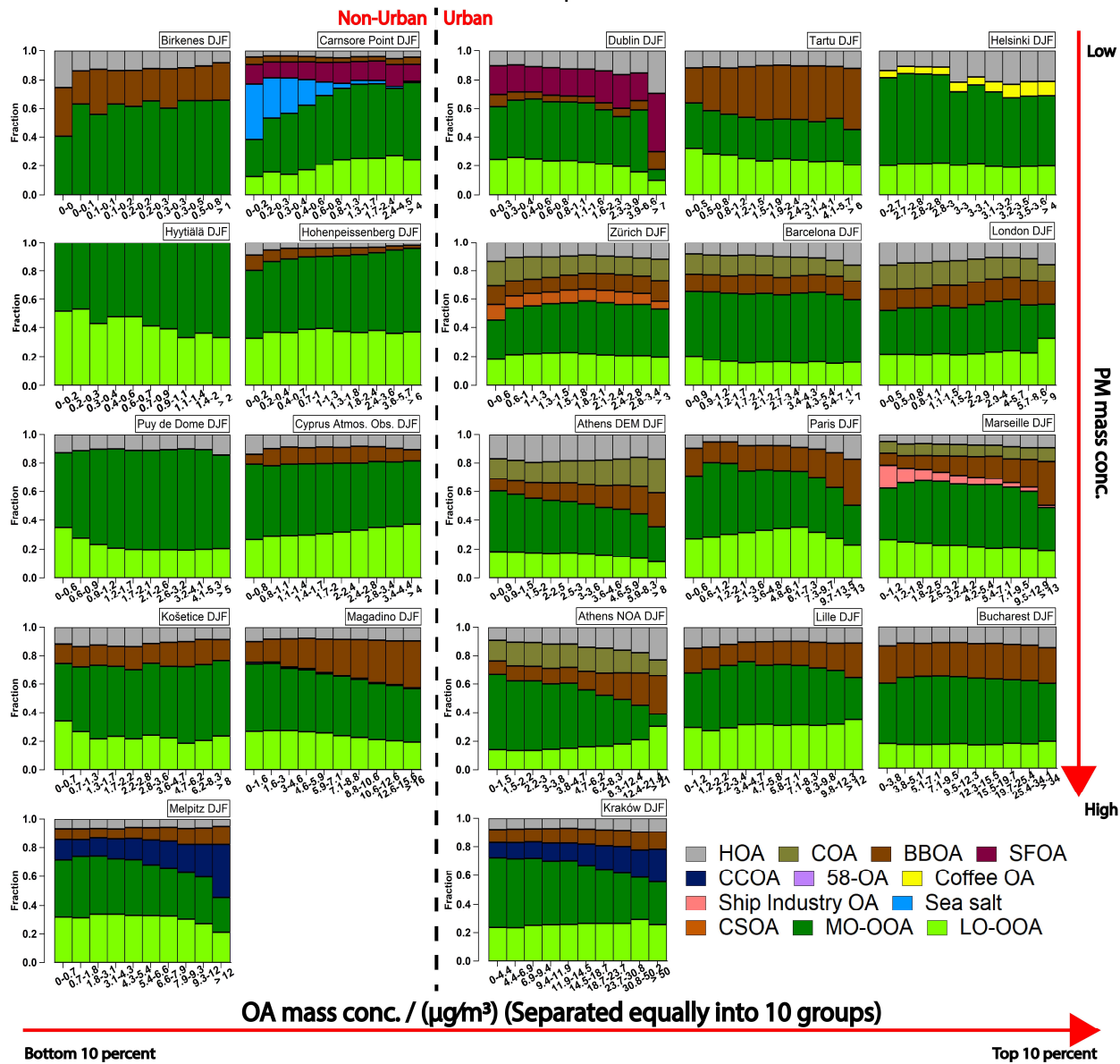

**Fig. S2.** Mass fractions of the organic aerosol (OA) components in winter (December, January and February; DJF) for 22 stations with 10 equally distributed bins (based on OA mass concentration). Non-urban and urban sites have been divided into left and right panels. The stations are sorted based on the total PM mass concentration (PM mass conc.).

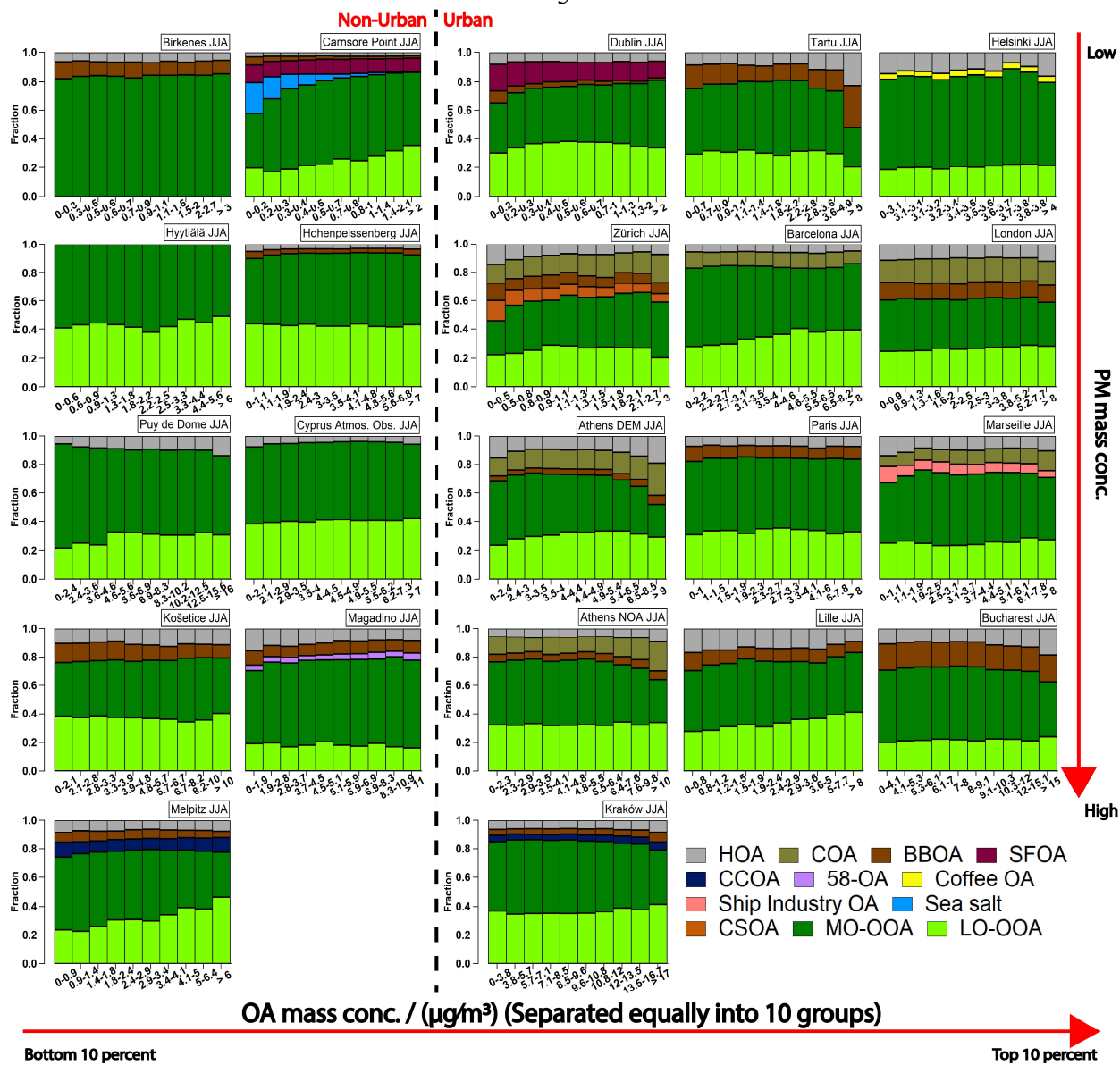

**Fig. S3** Mass fractions of the organic aerosol (OA) components in summer (June, July and August, JJA) for 22 stations with 10 equally distributed bins (based on OA mass concentration). Non-urban and urban sites have been divided into left and right panels. The stations are sorted based on the total PM mass concentration (PM mass conc.).

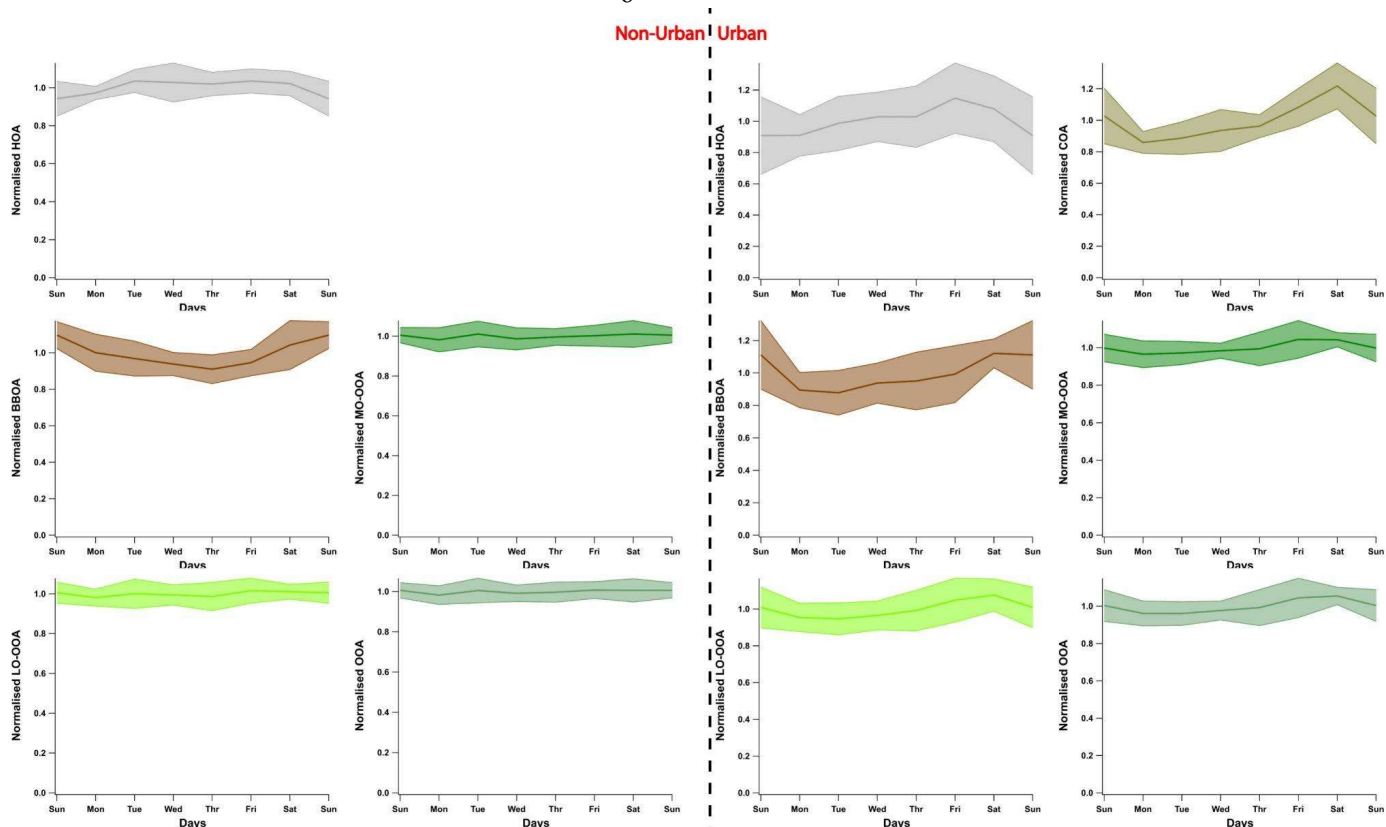

**Fig. S4** Normalised weekly cycles for HOA, BBOA, COA, MO-OOA, LO-OOA, and total OOA (MO-OOA+LO-OOA). The solid line represents the average of the normalised weekly cycles over urban/non-urban datasets, the shaded areas represent the corresponding standard deviation. In general, the primary organic aerosol (POA) factors show more pronounced weekly cycles than the OOA factors. In addition, urban sites have stronger weekly patterns than non-urban sites.

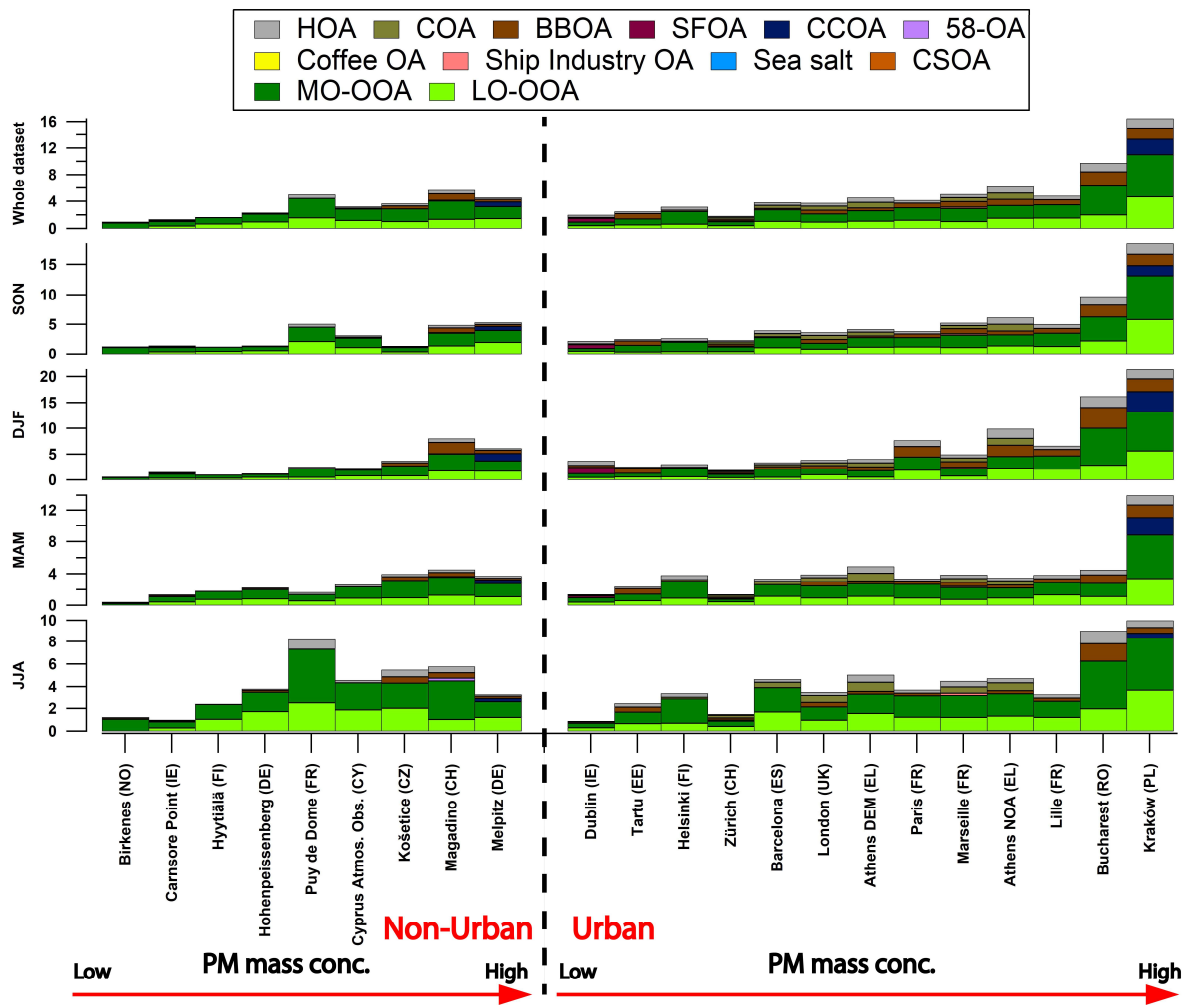

**Fig. S5.** Absolute mass concentrations (in µg/m$^3$) of all OA components at each station grouped by season (see text). The stations are categorised into non-urban and urban sites and by ascending order from low to high PM$_1$ mass concentrations (PM mass conc.).

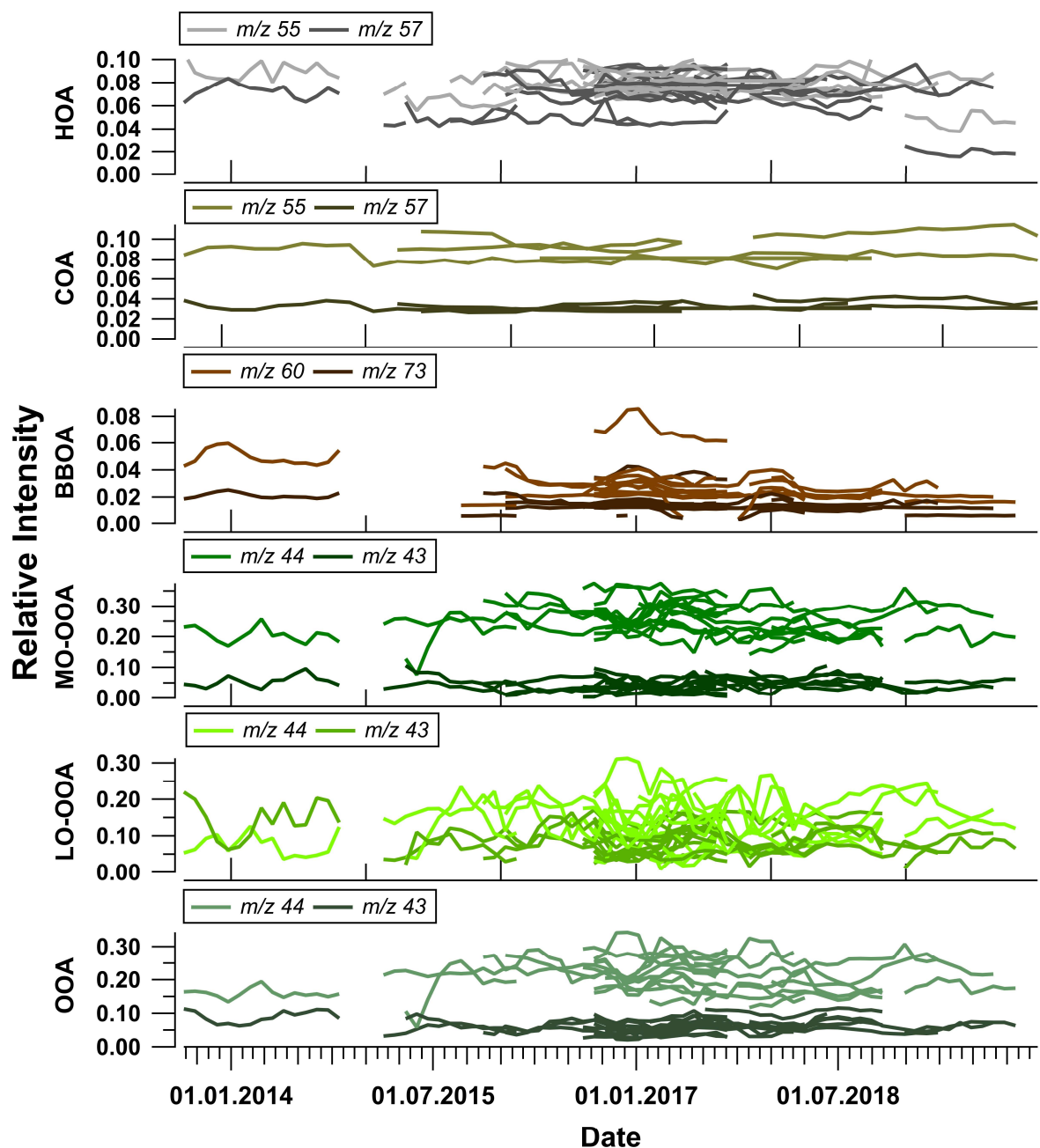

**Fig. S6.** Time series (monthly averaged) of key ions for the resolved OA factors (i.e., *m/z* 55 and *m/z* 57 for HOA and COA; *m/z* 60 and *m/z* 73 for BBOA; *m/z* 44 and *m/z* 43 for MO-OOA, LO-OOA, and Total OOA).

**Table S1.** Description of Each Dataset.

| Location | Acronym | CE[1] | Type | Link | Publication |
|---|---|---|---|---|---|
| Birkenes (NO) | bir | 0.45 | non-urban | http://ebas-data.nilu.no/ | (Yttri et al., 2021) |
| Carnsore Point (IE) | casp | 1 | non-urban | www.macehead.org | (Lin et al., 2019) |
| Hyytiälä (FI) | hyy | Other[2] | non-urban | N/A | (Heikkinen et al., 2020) |
| Hohenpeissenberg (DE) | hoh | 0.50 | non-urban | https://www.dwd.de/EN/research/observing_atmosphere/composition_atmosphere/hohenpeissenberg/cont_nav/start_mohp.html | N/A |
| Puy de Dôme (FR) | pdd | 0.50 | non-urban | N/A | N/A |
| Cyprus Atmos. Obs. (CY) | cao | CDCE[3] | non-urban | https://cao.cyi.ac.cy/ | N/A |
| Košetice (CZ) | kos | CDCE | non-urban | https://www.actris-ri.cz/en/menu/naok/ | N/A |
| Magadino (CH) | mag | 0.45 | non-urban | https://www.empa.ch/web/s503/nabel | (Chen et al., 2021) |
| Melpitz (DE) | mel | CDCE | non-urban | https://www.tropos.de/en/research/projects-infrastructures-technology/coordinated-observations-and-networks/tropos-research-site-melpitz | (Paglione et al., 2020) |
| Dublin (IE) | dub | 1 | urban | www.macehead.org | (Lin et al., 2019) |
| Tartu (EE) | tar | CDCE | urban | N/A | N/A |
| Helsinki (FI) | hel | CDCE | urban | https://julkaisu.hsy.fi/en/index/ilmanlaadun-mittausasemat-vuonna-2021.html#cLQTCX3SrU | (Barreira et al., 2021) |
| Athens DEM (EL) | athd | 0.5 | urban | N/A | N/A |
| Zürich (CH) | zur | 1 | urban | https://www.empa.ch/web/s503/nabel | N/A |
| Barcelona (ES) | bar | CDCE | urban | http://ebas-data.nilu.no/ | (Via et al., 2021) |
| London (UK) | lon | 0.45 | urban | N/A | N/A |
| Paris (FR) | par | CDCE | urban | https://sirta.ipsl.fr/ | (Petit et al., 2021; Zhang et al., 2019) |
| Marseille (FR) | mar | CDCE | urban | https://www.hermes-aq.com/ | (Chazeau et al., 2022, 2021); |
| Athens NOA (EL) | athn | CDCE | urban | N/A | (Stavroulas et al., 2019) |
| Lille (FR) | lil | CDCE | urban | http://ebas-data.nilu.no/ | Chebaicheb et al. (in prep.) |
| Bucharest (RO) | buc | 0.5 | urban | http://ebas-data.nilu.no/ | N/A |
| Kraków (PL) | kra | 0.5 | urban | N/A | (Tobler et al., 2021) |

[1] Collection Efficiency
[2] Differential Mobility Particle Sizer (DMPS)-based CE correction (Heikkinen et al., 2020)
3 Composition Dependent Collection Efficiency (Middlebrook et al., 2012)

Table S2. List of suggested criteria for the harmonised source apportionment of the 22 datasets.

| # | Criterion | Type | Threshold | Comments |
|---|---|---|---|---|
| 1 | HOA vs $NO_x$ | $R^2$, normal time series | p-value ≤ 0.05 | Need to think about the potential sources of NOx to consider which is the better criterion to validate HOA |
| 2 | HOA vs $eBC_{ff}$ | $R^2$, normal time series | p-value ≤ 0.05 | |
| 3 | BBOA vs $eBC_{wb}$ | $R^2$, normal time series | p-value ≤ 0.05 | Use criterion #4 instead if there are too many missing points in winter |
| 4 | Explained Variation [60] by BBOA | Average, normal time series | to-factor (p-value ≤ 0.05) | Investigate the Explained variation of m/z 60 by BBOA, make sure it explained most of 60 variabilities by this fresh BBOA |
| 5 | Explained Variation [115] by CCOA | Average, normal time series | to-factor (p-value ≤ 0.05) | Investigate the Explained variation of m/z 115 by BBOA, make sure it explained most of 115 variabilities by this fresh CCOA |
| 6 | COA[12]/(COA[9]+ COA[10])/2 | Average, hours | >1 | Inspect the exact hours of lunch peak in the seasonal solutions, then change 12PM in the numerator in the equation correspondingly |
| 7 | [4]factor_4[44] | Profiles, fraction, Sorting criterion | >0 | Sort LV/MO-OOA based on the $f_{44}$ intensity for the first unconstrained factor |
| 8 | [5]factor_5[43] | Profiles, fraction | >0 | Monitoring intensity of $f_{43}$ for the second unconstrained factor to make sure it is larger than 0 |
| 9 | factor_5[60] | Profiles, fraction | Take all | OPTIONAL, for monitoring purpose, just to check the mixing between BBOA and SV/LO-OOA |
| 10 | factor_4[60] | Profiles, fraction | Take all | OPTIONAL, for monitoring purpose, just to check the mixing between BBOA and LV/MO-OOA |

---

[4] factor_4 refers to the first un-constrained factor
[5] factor_5 refers to the second un-constrained factor

**Table S3.** Correlation ($R^2_{Pearson}$) between major OA factors and corresponding external data ($p < 0.05$).

| $R^2$ | HOA vs. eBC | BBOA vs. Org60 | MO-OOA vs. $SO_4$ | LO-OOA vs. $NO_3$ | Total OOA vs. $NH_4$ |
|---|---|---|---|---|---|
| **Birkenes (NO)** | 0.33 | 0.81 | | N/A | 0.17 |
| **Carnsore Point (IE)** | 0.48 | 0.65 | 0.38 | 0.44 | 0.72 |
| **Hyytiälä (FI)** | N/A | N/A | 0.22 | 0.11 | 0.12 |
| **Hohenpeissenberg (DE)** | 0.15 | 0.97 | 0.60 | 0.24 | 0.40 |
| **Puy de Dôme (FR)** | 0.39 | N/A | 0.0.02 | 0.01 | N/A |
| **Cyprus Atmos. Obs. (CY)** | 0.24 | 0.54 | 0.51 | 0.34 | 0.50 |
| **Kosetice (CZ)** | 0.12 | 0.84 | 0.44 | 0.01 | 0.20 |
| **Magadino (CH)** | 0.51 | 0.95 | 0.49 | 0.32 | 0.44 |
| **Melpitz (DE)** | 0.57 | 0.92 | 0.64 | 0.42 | 0.57 |
| **Dublin (IE)** | 0.71 | 0.96 | 0.41 | 0.29 | 0.60 |
| **Tartu (EE)** | 0.50 | 0.97 | 0.27 | 0.40 | 0.45 |
| **Helsinki (FI)** | 0.43 | N/A | 0.22 | 0.22 | 0.22 |
| **Athens DEM (EL)** | 0.47 | 0.92 | 0.42 | 0.06 | 0.46 |
| **Zürich (CH)** | 0.25 | 0.56 | 0.37 | 0.06 | 0.19 |
| **Barcelona (ES)** | 0.67 | 0.87 | 0.18 | 0.07 | 0.41 |
| **London (UK)** | 0.79 | 0.86 | 0.19 | 0.12 | 0.19 |
| **Paris (FR)** | 0.87 | 0.95 | 0.54 | 0.67 | 0.73 |
| **Marseille (FR)** | 0.4 | 0.9 | 0.2 | 0.15 | 0.31 |
| **Athens NOA (EL)** | 0.59 | 0.91 | 0.34 | 0.43 | 0.38 |
| **Lille (FR)** | 0.61 | 0.94 | 0.46 | 0.54 | 0.69 |
| **Bucharest (RO)** | 0.29 | 0.82 | 0.54 | 0.37 | 0.8 |
| **Kraków (PL)** | 0.78 | 0.92 | 0.22 | 0.46 | 0.50 |

**Table S4.** PMF errors (std/mean conc.) of major OA factors (estimated by logarithmic probability density functions (pdf) of the standard deviations of each time point i divided by the mean concentration of each time point i for corresponding OA factors, Equation (6) in Canonaco et al. (2021)).

| *PMF Errors (%)* | *HOA* | *COA* | *BBOA* | *MO-OOA* | *LO-OOA* | *Total OOA* |
|---|---|---|---|---|---|---|
| **Birkenes (NO)** | 31.9 | N/A | 34.0 | 4.5 | N/A | 4.5 |
| **Carnsore Point (IE)** | 29.6 | N/A | 32.8 | 12.9 | 33.3 | 21.3 |
| **Hyytiälä (FI)** | N/A | N/A | N/A | 20.7 | 30.6 | |
| **Hohenpeissenberg (DE)** | 44.3 | N/A | 36.9 | 33.1 | 44.2 | 20.2 |
| **Puy de Dôme (FR)** | 26.4 | N/A | N/A | 38.9 | 17.3 | 23.7 |
| **Cyprus Atmos. Obs. (CY)** | 11.7 | N/A | 6.5 | 8.3 | 11.5 | 8.1 |
| **Košetice (CZ)** | 38.2 | N/A | 27.0 | 22.2 | 40.5 | 21.0 |
| **Magadino (CH)** | 16.9 | N/A | 14.3 | 19.4 | 34.4 | 19.8 |
| **Melpitz (DE)** | 32.5 | N/A | 27.2 | 24.4 | 38.1 | 22 |
| **Dublin (IE)** | 18.4 | N/A | 28.6 | 49.2 | 58.1 | 32.9 |
| **Tartu (EE)** | 11.6 | N/A | 7.1 | 24.2 | 44.4 | 28.2 |
| **Helsinki (FI)** | 9.1 | N/A | N/A | 12.2 | 35.6 | 30.2 |
| **Athens DEM (EL)** | 18.7 | 26.4 | 16.8 | 24.5 | 39.6 | 22.3 |
| **Zürich (CH)** | 21.0 | 15.7 | 26.5 | 14.6 | 28.3 | 18.7 |
| **Barcelona (ES)** | 12.8 | 10.4 | 18.2 | 21.4 | 42.8 | 27.2 |
| **London (UK)** | 19.6 | 23 | 20.7 | 39.4 | 75.0 | 36.5 |
| **Paris (FR)** | 17.7 | N/A | 15.6 | 26.7 | 40 | 19.5 |
| **Marseille (FR)** | 21.1 | 24.2 | 31.1 | 34.1 | 15.2 | 21.2 |
| **Athens NOA (EL)** | 19.3 | 13.8 | 23.4 | 11.7 | 26.6 | 18.5 |
| **Lille (FR)** | 15.8 | N/A | 15.2 | 26.8 | 39.1 | 20.3 |
| **Bucharest (RO)** | 21.7 | N/A | 28.6 | 22 | 44.4 | 25.1 |
| **Kraków (PL)** | 27.1 | N/A | 26.1 | 21.8 | 39.2 | 22.0 |
| *Average Errors* | 20.0 ± 10.1 | 16.4 ± 6.3 | 23.0 ± 8.7 | 23.3 ± 10.8 | 37.1 ± 13.9 | 22.1 ± 7.1 |

Table S5. Contribution (in %) of major OA components for each season.

| Contributions (%) | HOA | | | | | BBOA | | | | | COA | | | | | MO-OOA | | | | | LO-OOA | | | | | Other OA | | | | |
|---|---|---|---|---|---|---|---|---|---|---|---|---|---|---|---|---|---|---|---|---|---|---|---|---|---|---|---|---|---|---|
| Seasons | All | SON | DJF | MAM | JJA | All | SON | DJF | MAM | JJA | All | SON | DJF | MAM | JJA | All | SON | DJF | MAM | JJA | All | SON | DJF | MAM | JJA | All | SON | DJF | MAM | JJA |
| Birkenes (NO) | 6.0 | 3.3 | 7.0 | 20.0 | 5.6 | 11.0 | 7.3 | 12.3 | 31.1 | 8.9 | | | | | | 83.0 | 89.4 | 80.7 | 48.9 | 85.5 | | | | | | | | | | |
| Carnsore Point (IE) | 3.7 | 4.2 | 4.6 | 3.6 | 2.0 | 3.0 | 2.8 | 3.9 | 2.2 | 2.0 | | | | | | 50.0 | 49.0 | 50.7 | 49.6 | 54.1 | 28.4 | 28.7 | 24.3 | 31.4 | 29.6 | 14.9 | 15.4 | 16.4 | 13.1 | 12.2 |
| Hyytiälä (FI) | | | | | | | | | | | | | | | | 57.8 | 60.2 | 62.5 | 57.5 | 55.5 | 42.2 | 39.8 | 37.5 | 42.5 | 44.5 | | | | | |
| Hohenpeissenberg (DE) | 4.2 | 5.5 | 4.8 | 4.4 | 3.4 | 5.5 | 6.9 | 6.4 | 6.6 | 4.5 | | | | | | 48.1 | 46.9 | 45.6 | 52.0 | 46.9 | 42.2 | 40.7 | 43.2 | 37.0 | 45.3 | | | | | |
| Puy de Dôme (FR) | 11.5 | 11.0 | 12.2 | 18.8 | 10.7 | | | | | | | | | | | 56.7 | 47.5 | 67.2 | 47.1 | 58.5 | 31.7 | 41.6 | 20.6 | 34.1 | 30.7 | | | | | |
| Cyprus Atmos. Obs. (CY) | 8.2 | 8.9 | 10.0 | 11.2 | 5.2 | 3.7 | 6.3 | 9.6 | | | | | | | | 51.2 | 49.5 | 46.7 | 54.3 | 53.8 | 36.9 | 35.2 | 33.7 | 34.5 | 41.0 | | | | | |
| Košetice (CZ) | 9.7 | 10.7 | 10.4 | 7.8 | 11.3 | 12.1 | 9.3 | 15.5 | 11.9 | 10.2 | | | | | | 50.3 | 51.4 | 51.9 | 55.4 | 41.5 | 28.0 | 28.6 | 22.2 | 24.8 | 37.1 | | | | | |
| Magadino (CH) | 9.3 | 9.8 | 9.3 | 8.1 | 9.3 | 17.1 | 16.6 | 27.4 | 11.2 | 8.3 | | | | | | 47.2 | 43.4 | 40.2 | 49.8 | 60.2 | 24.3 | 28.2 | 22.3 | 28.6 | 18.0 | 2.1 | 2.0 | 0.7 | 2.2 | 4.2 |
| Melpitz (DE) | 6.5 | 6.6 | 5.9 | 7.3 | 6.9 | 7.8 | 6.2 | 10.6 | 6.5 | 6.0 | | | | | | 38.0 | 37.7 | 32.2 | 44.6 | 41.7 | 32.4 | 37.0 | 28.3 | 30.1 | 36.6 | 15.2 | 12.6 | 23.1 | 11.6 | 8.7 |
| Dublin (IE) | 18.7 | 21.6 | 23.8 | 9.9 | 6.6 | 7.4 | 6.8 | 10.2 | 4.3 | 2.2 | | | | | | 26.1 | 21.2 | 19.5 | 39.7 | 42.9 | 21.2 | 22.1 | 13.6 | 29.1 | 35.2 | 26.6 | 28.4 | 32.9 | 17.0 | 13.2 |
| Tartu (EE) | 11.5 | 12.3 | 11.0 | 10.9 | 13.8 | 31.3 | 29.4 | 38.3 | 29.0 | 18.2 | | | | | | 35.7 | 43.3 | 27.3 | 34.9 | 40.5 | 21.4 | 15.1 | 23.5 | 25.2 | 27.5 | | | | | |
| Helsinki (FI) | 15.6 | 18.2 | 17.9 | 14.4 | 12.5 | | | | | | | | | | | 58.3 | 59.5 | 54.1 | 56.5 | 62.7 | 20.6 | 15.9 | 20.5 | 24.3 | 20.7 | 5.6 | 6.4 | 7.5 | 4.7 | 4.1 |
| Athens DEM (EL) | 15.4 | 11.7 | 17.7 | 17.6 | 13.2 | 9.0 | 6.7 | 17.7 | 7.1 | 4.7 | 17.8 | 15.5 | 19.4 | 19.4 | 15.8 | 34.0 | 38.4 | 31.0 | 32.4 | 35.2 | 23.9 | 27.7 | 14.3 | 23.5 | 31.0 | | | | | |
| Zürich (CH) | 9.7 | 11.5 | 10.2 | 8.5 | 7.2 | 13.4 | 14.1 | 14.8 | 11.3 | 11.2 | 15.6 | 16.2 | 13.3 | 17.6 | 15.8 | 30.6 | 32.1 | 33.2 | 23.2 | 30.9 | 24.2 | 18.8 | 23.0 | 32.4 | 28.3 | 6.5 | 7.3 | 5.6 | 7.0 | 6.6 |
| Barcelona (ES) | 11.4 | 13.5 | 12.2 | 9.7 | 6.0 | 4.7 | 4.3 | 12.8 | 0.3 | | 12.1 | 13.0 | 12.2 | 12.4 | 10.1 | 44.2 | 41.9 | 46.8 | 43.7 | 47.3 | 27.6 | 27.3 | 16.0 | 33.9 | 36.6 | | | | | |
| London (UK) | 12.4 | 14.2 | 13.6 | 10.2 | 10.7 | 14.8 | 17.7 | 15.9 | 13.3 | 11.3 | 15.6 | 18.8 | 14.1 | 13.3 | 17.2 | 32.0 | 26.4 | 29.5 | 39.3 | 33.1 | 25.1 | 22.9 | 26.9 | 24.0 | 27.7 | | | | | |
| Paris (FR) | 10.5 | 12.2 | 15.5 | 8.5 | 7.8 | 16.6 | 16.1 | 26.5 | 12.6 | 8.3 | | | | | | 43.5 | 40.5 | 33.1 | 50.9 | 50.7 | 29.5 | 31.2 | 24.9 | 27.9 | 33.2 | | | | | |
| Marseille (FR) | 10.6 | 8.7 | 14.1 | 12.1 | 11.9 | 15.9 | 16.1 | 22.1 | 13.4 | | 11.0 | 10.2 | 13.9 | 12.1 | 11.9 | 37.7 | 39.3 | 30.1 | 38.7 | 42.7 | 20.4 | 20.9 | 15.5 | 19.8 | 27.2 | 4.3 | 4.7 | 4.4 | 3.9 | 6.2 |
| Athens NOA (FR) | 15.3 | 18.2 | 18.6 | 10.6 | 8.4 | 14.5 | 10.4 | 21.9 | 10.6 | 5.5 | 15.2 | 18.7 | 13.4 | 15.4 | 14.7 | 30.3 | 29.9 | 22.9 | 36.3 | 43.2 | 24.7 | 22.8 | 23.3 | 27.1 | 28.2 | | | | | |
| Lille (FR) | 12.2 | 14.5 | 11.1 | 12.5 | 12.3 | 15.1 | 15.9 | 18.6 | 12.2 | 8.4 | | | | | | 40.3 | 43.8 | 38.2 | 40.5 | 42.3 | 32.4 | 25.8 | 32.1 | 34.8 | 36.9 | | | | | |
| Bucharest (RO) | 13.7 | 15.2 | 13.3 | 14.4 | 12.8 | 21.9 | 20.6 | 24.7 | 23.7 | 17.3 | | | | | | 43.5 | 41.5 | 44.2 | 36.6 | 47.7 | 20.9 | 22.7 | 17.8 | 25.3 | 22.2 | | | | | |
| Kraków (PL) | 8.6 | 9.7 | 8.7 | 8.8 | 6.5 | 10.4 | 10.2 | 11.5 | 11.6 | 5.2 | | | | | | 38.3 | 39.4 | 35.2 | 39.2 | 46.5 | 28.5 | 31.5 | 26.4 | 24.5 | 37.3 | 14.1 | 9.1 | 18.2 | 16.0 | 4.5 |